\documentclass[letter]{IEEEtran}

\usepackage[utf8]{inputenc} 
\usepackage[T1]{fontenc}
\usepackage{url}
\usepackage{ifthen}
\usepackage{cite}
\usepackage[cmex10]{amsmath} 
\usepackage{dsfont}
\usepackage{booktabs}

\ifCLASSINFOpdf
  \usepackage[pdftex]{graphicx}
\else
  \usepackage[dvips]{graphicx}
\fi
\usepackage{kotex}
\usepackage{amsfonts, amssymb}
\usepackage{xcolor}
\usepackage{stmaryrd}

\usepackage{lipsum}

\begin{document}

\title{Learning While Transmitting: Pilotless Polar Coded Modulation for Short Packet Transmission}

\author{   
Geon Choi, \IEEEmembership{Member, IEEE} and
 Namyoon Lee, \IEEEmembership{Senior Member, IEEE}
\thanks{The authors are with the Department of Electrical Engineering, Pohang University of Science and Technology (POSTECH), Pohang 37673, South Korea (e-mail: \mbox{geon.choi@postech.ac.kr}; \mbox{nylee@postech.ac.kr}).
}
}

\maketitle

\begin{abstract}
 Short packets make channel learning expensive. In pilot-aided transmission (PAT), a non-negligible fraction of the packet is consumed by pilots, creating a direct pre-log loss and tightening the reliability margin needed for ultra-reliable low-latency communication. We propose a pilot-free polar-coded framework that replaces explicit pilots with \emph{coded pilots}. The message is carried by two polar-coded segments: a quadrature phase shift keying
(QPSK) segment that is decodable without channel state information (CSI), and a higher-order quadrature amplitude modulation (QAM) segment that provides high spectral efficiency. The receiver employs \emph{hybrid decoding}: it first jointly infers CSI during successive-cancellation-based decoding of the QPSK segment by exploiting QPSK phase-rotation invariance together with polar frozen-bit constraints; the decoded QPSK symbols then act as \emph{implicit pilots} for coherent detection and decoding of the QAM segment. The split also makes rate adaptation practical by confining the symmetry/frozen-bit requirements for phase resolution to the QPSK segment, enabling puncturing and shortening without breaking the pilot-free mechanism. For multi-block fading, we optimize the split and code parameters via density evolution with Gaussian approximation (DEGA); for higher-order modulation, we use bit-interleaved coded modulation capacity approximation to obtain equivalent channel parameters. Incorporating channel-estimation error variance into the DEGA-based analysis, simulations over practical multi-block block-fading channels show gains up to $1.5$~dB over PAT in the short-blocklength regime.

 \end{abstract}

\section{Introduction}

Ultra-reliable low-latency communication (URLLC) changes the operating point of wireless design. When the packet is short and the error probability target is stringent, there is no longer room to \textit{average out} inefficiencies: every channel use spent on overhead directly reduces the information that can be delivered within the latency budget. This regime is increasingly central in emerging closed-loop applications---robot control for physical artificial intelligence, autonomous driving, and industrial automation---where the control cycle itself allocates only a small number of channel uses per update.

Polar codes, introduced by Ar{\i}kan~\cite{arikan09-polar}, have become a leading coding solution for this short-block, high-reliability regime. They are the first family of channel codes proven to achieve the capacity of symmetric binary-input memoryless channels under low-complexity successive cancellation (SC) decoding, and their structure supports efficient list decoding and practical implementations. These advantages led to their adoption in the 5G New Radio (NR) standard~\cite{3gpp-nr-coding}. Beyond standardization, a large body of work has improved finite-length performance through pre-transform and concatenation ideas, including cyclic redundancy check (CRC)-aided decoding, dynamic frozen bits, polarization-adjusted convolutional (PAC) code-type constructions, and learning-assisted methods~\cite{Niu12-CA-polar, Trifonov-polar-dynamic-frozen, Wang-PCC-polar, arikan-pac, deep-polar-tcom, spp-tcom, Rate-matching-deep-polar}. In short, from a \emph{coding} perspective, the ingredients for URLLC are strong~\cite{shirvanimoghaddam19, yue23, stephan-welcome-chance, BOSS-URLLC, BOSS3}.

The more subtle bottleneck is often not coding, but \emph{coherence}. Consider the basic block-fading picture: within a coherence interval of $T$ channel uses, the channel is (approximately) constant but unknown. Pilot-assisted transmission (PAT) spends $\tau$ of these $T$ uses on pilots, leaving only $T-\tau$ uses for data~\cite{URLLC2018, Tong04-PAT-mag, Wang-6G-Survey, david-6G-URLLC}. The overhead creates a fundamental pre-log penalty: even before accounting for estimation error, the effective payload is reduced by a factor $(1-\tau/T)$. When $T$ is small---as in fast fading, high mobility, or short packets---this penalty is no longer negligible. Moreover, in the very regime where reliability must be highest, imperfect channel estimates can turn the remaining data symbols into a mismatch problem, compounding the loss. This motivates pilot-free (or pilot-minimized) designs that embed channel inference into the coded transmission itself, aiming to approach coherent performance without explicit pilot overhead~\cite{Shortpacket2016, Shortpacket2019, ZhengTse2002, BIM2021, PilotfreeMIMO2020, BOSSblind2025}.

\subsection{Related Works}

Pilot-free polar-coded communication has developed along several directions, each trading off generality, complexity, and robustness.

A first direction uses \emph{code constraints as implicit references}. Yuan \emph{et al.}~\cite{yuan21-polar-non-coherent} proposed a two-stage scheme that eliminates explicit pilots by exploiting frozen-bit constraints for joint channel estimation and decoding: candidate phase rotations are evaluated through successive cancellation list (SCL)  decoding and likelihood comparison. The approach achieves near-coherent performance for quadrature phase shift keying (QPSK), but scales poorly to multi-block fading because the number of phase combinations grows rapidly, and extending to higher-order modulation is challenging due to amplitude-related ambiguities.

A second direction resolves phase ambiguity by \emph{enforcing equivariance}. Phase-equivariant polar codes freeze rotation-discriminating bits to leverage code automorphisms and disambiguate unknown phase without pilots~\cite{stbrink23-phase-equivariant}. While effective for QPSK and 16-quadrature amplitude modulation (QAM), these constructions are fragile under the rate-matching operations required in practice: puncturing/shortening can disrupt frozen-bit patterns and symmetry, undermining phase resolution.

A third direction introduces \emph{embedded pilot positions}. Systematic polar codes with selected positions used for channel estimation were developed in~\cite{li18-systematic-polar-channel-est}, enabling operation in multi-carrier and Doppler channels. However, these reserved positions ultimately do not carry data, so overhead is not fully converted into coding gain. Hybrid polar encoding~\cite{zheng23-hybrid-polar} creates \textit{dynamic pilots} by mixing systematic and non-systematic segments, enabling blind estimation followed by coherent decoding; current formulations are mainly tailored to real-valued fading with binary phase shift keying (BPSK) and do not directly address unknown phase offsets in complex baseband channels.

These works reveal a recurring design tension: a practical pilot-free scheme must (i) support higher-order modulation while resolving phase ambiguities, (ii) remain compatible with flexible rate-matching and arbitrary code lengths, and (iii) keep complexity manageable in multi-block fading. Existing methods typically satisfy only a subset of these requirements.

\subsection{Overview of Approach}

We propose a pilot-free polar-coded transmission framework that resolves the above tension by separating two roles that are usually coupled: \emph{channel anchoring} and \emph{spectral efficiency}. The core idea is a \emph{code-splitting} architecture: a QPSK-modulated segment provides a constant-amplitude anchor that enables blind channel inference using the algebraic structure of polar codes, while a higher-order QAM segment carries the bulk of the information at high spectral efficiency. The receiver first decodes the QPSK segment to recover the relevant channel state information (CSI), then reuses the decoded bits as implicit pilots to enable coherent detection and decoding of the QAM segment. Importantly, by confining the phase-resolution constraints to the QPSK portion, the overall design remains compatible with practical rate-matching.

\subsection{Contributions}

Our main contributions are summarized as follows:
\begin{itemize}
\item \textbf{Code-splitting architecture:} We develop a pilot-free polar-coded framework that partitions each codeword into a QPSK segment for blind channel estimation and a higher-order QAM segment for spectral efficiency, supporting arbitrary block lengths.

\item \textbf{Hybrid decoding with implicit pilots:} We design an efficient two-phase receiver that first resolves CSI through decoding of the QPSK segment, then uses the recovered bits as implicit pilots for coherent QAM detection and decoding.

\item \textbf{Rate-matching integration:} The split localizes the delicate frozen-bit/symmetry constraints to the QPSK component, enabling puncturing/shortening (e.g., quasi-uniform rate-matching) without breaking phase-ambiguity resolution.

\item \textbf{Multi-block fading optimization:} We provide a systematic optimization framework for multiple fading blocks by extending density evolution with Gaussian approximation (DEGA) to the code-splitting setting. For higher-order modulation, we construct a binary input–additive white Gaussian noise (BI-AWGN) approximation channel and apply a bit-interleaved coded modulation (BICM) capacity-matching method to determine equivalent channel parameters.

\item \textbf{Analysis and validation:} We incorporate channel-estimation error variance into the DEGA-based analysis and validate the model through extensive simulations across diverse fading profiles and modulation orders, demonstrating gains up to $1.5$~dB over PAT in practical multi-block block-fading channels.
\end{itemize}

The remainder of this paper is organized as follows. Section~\ref{sec:prelim} presents the channel model and polar preliminaries. Section~\ref{sec:encoding} describes the proposed code-splitting architecture and encoding procedure. Section~\ref{sec:decoding} develops the hybrid decoding algorithm with implicit pilot generation. Section~\ref{sec:optimization} presents the DEGA- and BICM-based optimization framework. Section~\ref{sec:simulation} reports simulation results. Section~\ref{sec:conclusion} concludes the paper.

\section{System Model and Background}\label{sec:prelim}
\subsection{System Model}

We consider a packet of length $N_t$ transmitted over a block-fading channel with coherence time $N_c$. The packet experiences $B$ independent fading blocks, where $B = N_t / N_c$. The channel coefficient for the $b$th fading block is denoted as $h_b = |h_b|e^{j\phi_b}$.

Let the transmitted packet be represented as $\mathbf{x} = [\mathbf{x}_1, \mathbf{x}_2, \ldots, \mathbf{x}_B]$, where $\mathbf{x}_b \in \mathbb{C}^{N_c}$ for $b \in \{1, 2, \ldots, B\}$. The corresponding received signal vector $\mathbf{y} = [\mathbf{y}_1, \mathbf{y}_2, \ldots, \mathbf{y}_B]$ is given by
\begin{align}
\mathbf{y}_b = h_b\mathbf{x}_b + \mathbf{v}_b, \label{eq:channel_model}
\end{align}
where $\mathbf{v}_b \in \mathbb{C}^{N_c}$ denotes the additive white Gaussian noise (AWGN) vector with independent and identically distributed elements following $\mathbf{v}_b \sim \mathcal{CN}(\mathbf{0}, \sigma^2\mathbf{I})$.

Throughout this paper, we denote the $i$th element of vector $\mathbf{x}$ as $x_i$ and the $i$th element of vector $\mathbf{x}_k$ as $x_{k,i}$.


\subsection{Pilot-Aided   Communications}

 We begin by briefly reviewing the pilot-aided communication framework. In such systems, the transmitted symbol vector $\mathbf{x} \in \mathbb{C}^{N_t}$ is partitioned into $B+1$ components: pilot symbols $\mathbf{x}_p^{(b)} \in \mathbb{C}^{N_p^{(b)}}$ for the $b$th block fading, and data symbols $\mathbf{x}_d \in \mathbb{C}^{N_d}$, where $\sum_{b=1}^{B} N_p^{(b)} + N_d = N_t$. The pilot symbols, typically pseudo-random QPSK sequences of length $N_p^{(b)}$, are known a priori to the receiver and facilitate channel estimation. The data symbols, of length $N_d$, convey $K$ information bits. The pilot symbols are inserted between the data symbols, so that symbol vector $\mathbf{x}_p^{(b)}$ experiences the $b$th block fading channel.

Assuming $N_s$-ary QAM with constellation $\mathcal{X}_{N_s}$, each symbol $x_i \in \mathcal{X}_{N_s}$ represents $n_s = \log_2(N_s)$ consecutive codeword bits.\footnote{We assume the codeword bits are interleaved appropriately for the bit-interleaved coded modulation (BICM) scheme \cite{BICM-Carie-98}.} The constellation is normalized such that $\frac{1}{N_s}\sum_{x \in \mathcal{X}_{N_s}} |x|^2 = 1$, and Gray labeling is employed throughout \cite{3gpp-nr-modulation}.

Given a total packet size $N_c$, pilot length $N_p$, and modulation size $N_s$, the effective codeword length is $M = n_s N_d$. Accordingly, the transmitter encodes $K$ information bits $\mathbf{m} \in \mathbb{F}_2^K$ into a codeword $\mathbf{c} \in \mathbb{F}_2^M$ with nominal code rate $R = K /M$. However, due to the pilot overhead, the effective code rate becomes
\begin{align}
R_{\rm eff}^p = \frac{K}{N_c} = \frac{K}{N_p + N_d} = (1 - \alpha)n_sR,
\end{align}
where $\alpha = N_p / N_c$ denotes the pilot overhead fraction.




\subsection{Polar Codes and Rate-Matching}

Polar codes are defined by a transform matrix $\mathbf{F}_N = \mathbf{F}_2^{\otimes n}$ where $N = 2^n$ and $\mathbf{F}_2 = \begin{bmatrix} 1 & 0 \\ 1 & 1 \end{bmatrix}$. The code is characterized by an information index set $\mathcal{I} \subseteq \{0, 1, \ldots, N-1\}$, while the frozen index set $\mathcal{F}$ is the complement of $\mathcal{I}$.
The generator matrix consists of the rows of $\mathbf{F}_N$ indexed by $\mathcal{I}$. Polar codes achieve arbitrary code rates $R = K/N$ by setting $|\mathcal{I}| = K$.

The inherent constraint of Arıkan's construction limits polar code lengths to powers of two, i.e., $M = 2^n$. To support arbitrary codeword lengths $M \neq 2^n$, rate-matching techniques are employed:
\begin{itemize}
\item \textbf{Shortening and Puncturing:} Used to reduce code length from a larger mother code of size $N > M$. 
\item \textbf{Extension and Repetition:} Used to increase code length from a smaller mother code of size $N < M$.
\end{itemize}
In 5G NR systems, quasi-uniform puncturing and shortening with sub-block interleaving are employed for rate-matching, ensuring compatibility while maintaining good performance \cite{3gpp-nr-coding}. Shortening removes the last $N-M$ interleaved codeword bits, while puncturing removes the first $N-M$ bits. Repetition extends the first $M-N$ interleaved codeword bits.

\section{Pilot-Free Polar-Coded Modulation}\label{sec:encoding}

\begin{figure}
    \centering
    \includegraphics[width=\linewidth]{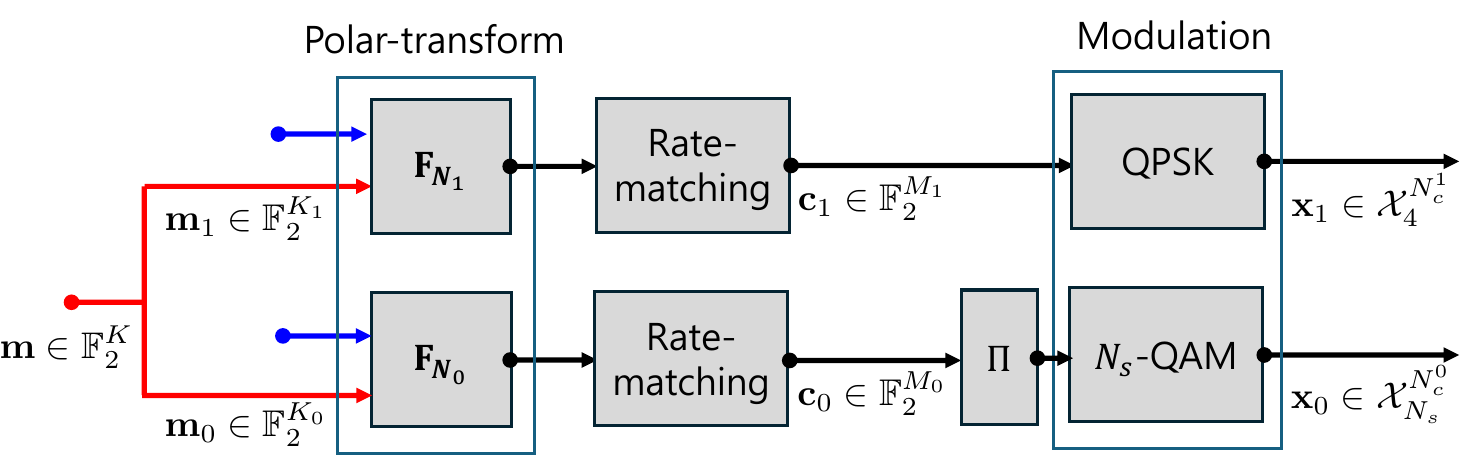}
    \caption{Encoding structure of the proposed code-splitting method. Information bits are partitioned into two sub-messages that are independently encoded and modulated using different schemes. $\Pi$ is an interleaver for BICM \cite{BICM-Carie-98}. }
    \label{fig:encoding}
\end{figure}

In this section, we present a rate-matching, pilot-free polar-coded transmission framework designed for efficient short-packet communication over fading channels. The core of our approach is a code-splitting architecture that separates the transmission into multiple SC-decodable\footnote{We use the term SC-decodable to indicate codes able to be decoded with SC types of decoder with low complexity} segments.

Consider the transmission of $K$ information bits over $N_t$ channel uses, comprised of $B$ independent fading blocks with coherence time $N_c$ symbols.
The proposed scheme is parameterized by $\{N_c^b, K_b\}_{b=0}^B$, satisfying the constraints $\sum_{b=0}^B N_c^b = N_t$ and $\sum_{b=0}^{B} K_b = K$. 
The overall encoding structure, illustrated in Fig.~\ref{fig:encoding}, follows a multi-path design in which the information message $\mathbf{m} \in \mathbb{F}_2^K$ is partitioned into $B+1$ sub-messages, $\mathbf{m}_b \in \mathbb{F}_2^{K_b}$, which are processed independently as follow:\footnote{The framework naturally accommodates outer codes such as CRC by treating the concatenated message $[\mathbf{m}, \mathbf{p}_{\text{crc}}]$ as the information sequence.}
\begin{itemize}
    \item \textbf{Component 0:} The sub-message $\mathbf{m}_0$ is encoded with a polar code $\mathcal{C}_0$, resulting in a codeword $\mathbf{c}_0$ with length $M_0 = N_c^0 \log2N_s$. We use conventional rate-matching techniques. Following interleaving, the codeword bits are modulated with $N_s$-QAM symbol vector $\mathbf{x}_0$ of length $N_c^0$. The symbol vector $\mathbf{x}_0$ is further divided into ${\bf x}_0 = [{\bf x}_{0, 1}, {\bf x}_{0,2}, \ldots, {\bf x}_{0, B}]$. 
    
    \item \textbf{Components $i$ ($i = 1, 2, \ldots, B$) -- coded-pilot:} Each sub-message $\mathbf{m}_i$ is encoded using polar code $\mathcal{C}_i$ with the constraint $\{N_i-2, N_i-1\} \subseteq \mathcal{F}_i$, producing codeword $\mathbf{c}_i$ of length $M_i = 2N_c^i$. We use conventional rate-matching techniques, but exclude shortening. The codeword is then QPSK-modulated to generate symbol vector $\mathbf{x}_i$ of length $N_c^i$, which serves as the channel estimation component for the $i$th fading block with channel coefficient $h_i$.
\end{itemize}

The complete transmitted signal consists of coded-pilot components $\mathbf{x}_i$ for channel estimation and higher-order modulated sub-vectors $\mathbf{x}_{0,i}$ for data transmission, arranged as $\mathbf{x} = [\mathbf{x}_1, \mathbf{x}_{0,1}, \mathbf{x}_2, \mathbf{x}_{0,2}, \ldots, \mathbf{x}_B, \mathbf{x}_{0,B}]$. Consequently, each sub-vector pair $[\mathbf{x}_b, \mathbf{x}_{0,b}]$ experiences the $b$th fading block with channel coefficient $h_b$.


The effective code rate of the proposed pilot-free polar-coded transmission scheme is given by
\begin{align}
    R_{\text{eff}} = \frac{K}{M} = n_s\alpha_0R_0 + \sum_{b=1}^B 2\alpha_bR_b,
\end{align}
where $M=\sum_{b=0}^{B} M_b$, $\alpha_b = N_c^b / N_t$ and $R_b = K_b / M_b$.
This effective code rate formulation naturally includes the pilot-aided case as a special instance. By selecting $K_b = 0$ and $N_c^{b} = N_p^{(b)}$ for $b\ge 1$, we recover the effective code rate for conventional pilot-aided transmission, i.e., $R_{\rm eff} = R_{\rm eff}^p$. Thus, the proposed framework generalizes the standard pilot-aided communication system. Furthermore, by appropriately choosing the parameters $(K_b, N_c^b)$ for $b \ge 0$, the effective code rate can be flexibly optimized to achieve a desired reliability target or spectral efficiency constraint.

In addition, the proposed method can achieve higher spectral efficiency compared to pilot-aided communication. When the decoding of ${\bf y}_{b \ge 1}$ is successful, the receiver can leverage the decoded information $\hat{\bf x}_{b \ge 1}$ for enhanced channel estimation. Since these successfully decoded symbols serve as known reference signals, they effectively function as pilot symbols while maintaining channel estimation quality equivalent to that of conventional pilot-based methods. However, unlike traditional pilot schemes where dedicated symbols carry no information, our approach enables ${\bf x}_{b \ge 1}$ to simultaneously transmit data bits and provide channel estimation capability.




\begin{figure}
    \centering
    \includegraphics[width=\linewidth]{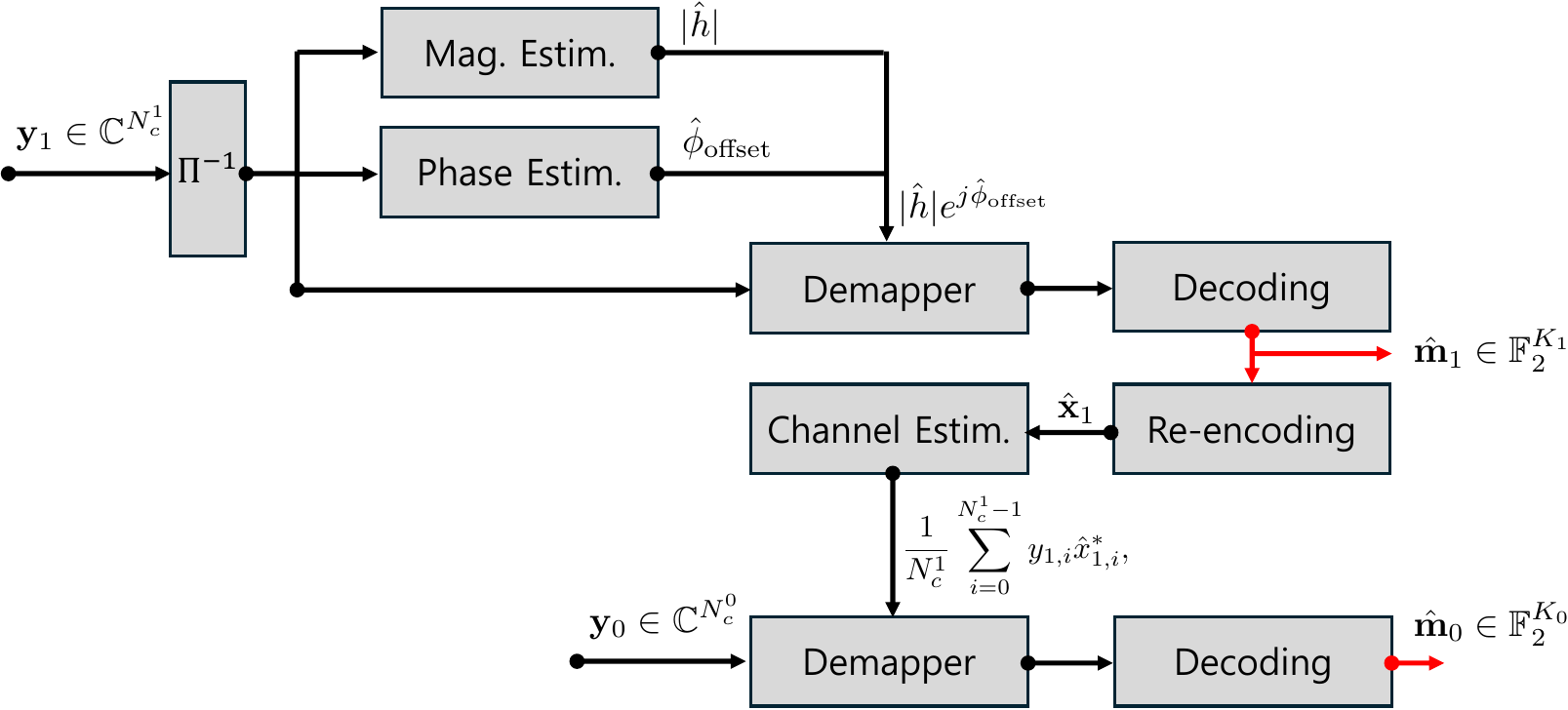}
    \caption{Decoding flowchart showing sequential processing of QPSK and QAM portions with blind channel estimation.}
    \label{fig:decoding}
\end{figure}

\section{Hybrid Decoding}\label{sec:decoding}

In this section, we present a hybrid decoding method. For simplicity, we restrict our attention to the single block-fading channel case with $B = 1$.
The decoding procedure operates in two stages. First, the coded-pilot component $\mathbf{y}_1$ is decoded in a blind fashion, yielding the decoded message $\hat{\mathbf{m}}_1$. This message is then re-encoded to generate a pilot sequence for estimating the channel coefficient $h_1$. Second, using the estimated channel, the higher-order modulated component $\mathbf{y}_0$ is decoded in a coherent manner.

\subsection{Blind Decoding for ${\bf x}_1$}
We now describe the decoding process for $\mathbf{m}_1$, which leverages both the constant-amplitude property of QPSK modulation and the algebraic structure of polar codes. As illustrated in Fig.~\ref{fig:decoding}, the proposed blind decoding procedure consists of three main stages:
(i) channel magnitude estimation, and
(ii) channel phase estimation.

\subsubsection{Channel Magnitude Estimation}

For constant-amplitude modulation such as QPSK where $|x_{1,i}|^2 = 1$, channel magnitude estimation leverages the relationship
\begin{align}
|y_{1,i}|^2 = |h|^2|x_{1,i}|^2 + |v_{1,i}|^2 + 2\textsf{Re}(h^* x_{1,i}^* v_{1,i}). \label{eq:magnitude_expansion}
\end{align}
Since the noise components $v_{1,i}$ are i.i.d. with zero mean, averaging over all received symbols yields
\begin{align}
\frac{1}{N_c^1}\sum_{i=0}^{N_c^1-1}|y_{1,i}|^2 \approx |h|^2 + \sigma^2, \label{eq:magnitude_average}
\end{align}
providing the magnitude estimate
\begin{align}
|\hat{h}| = \sqrt{\max\left(0, \frac{1}{N_c^1}\sum_{i=0}^{N_c^1-1}|y_{1,i}|^2 - \sigma^2 \right)}, \label{eq:magnitude_estimate}
\end{align}
where the $\max(\cdot, 0)$ operation ensures non-negative estimates in the presence of estimation errors.

\subsubsection{Channel Phase Estimation}

Channel phase estimation is more challenging due to the rotational symmetry inherent in QPSK modulation. The channel phase $e^{j\phi}$ rotates all received symbols by $\phi$ radians. However, the four-fold rotational symmetry of QPSK creates an inherent ambiguity in phase estimation, as rotations by multiples of $\pi/2$ cannot be distinguished without additional information.

Due to this phase ambiguity, the estimated phase $\hat{\phi}$ can be decomposed as
\begin{align}
\hat{\phi} = m\frac{\pi}{2} + \hat{\phi}_{\text{offset}}, \label{eq:phase_decomposition}
\end{align}
where $m \in \{0, 1, 2, 3\}$ represents integer ambiguity and $\hat{\phi}_{\text{offset}} \in [0, \pi/2)$ is the fractional offset.
We adopt the estimation strategy from \cite{stbrink23-phase-equivariant}, involving two steps: estimating the fractional offset without channel decoding, then resolving the integer ambiguity using the polar code structure.

{\bf Fractional Phase Offset Estimation:} For fractional offset estimation, the method employs the Viterbi and Viterbi phase estimation (VVPE) algorithm \cite{Viterbi-VVPE}. The VVPE algorithm is designed for M-PSK modulation and operates by rotating all received symbols to a common reference point, thereby eliminating the modulation-dependent phase variations.

For QPSK $(M=4)$, the key insight is that raising each symbol to the fourth power eliminates the data-dependent phase variations:
\begin{equation}
4\phi = 4m\frac{\pi}{2} + 4\phi_{\text{offset}} = 4\phi_{\text{offset}}, \label{eq:vvpe_principle}
\end{equation}
since $4m\pi/2 = 2m\pi$ is a multiple of $2\pi$.

The fractional phase offset is estimated as \cite{Rice-RRC}
\begin{align}
\hat{\phi}_{\text{offset}} = \frac{1}{4}\tan^{-1}\left(\frac{\sum_{i=0}^{N_c^1-1} \textsf{Im}\left(\frac{y_{1,i}^4}{|y_{1,i}|^{3}}\right)}{\sum_{i=0}^{N_c^1-1} \textsf{Re}\left(\frac{y_{1,i}^4}{|y_{1,i}|^{3}}\right)}\right) - \frac{\pi}{4}. \label{eq:vvpe_estimate}
\end{align}

{\bf Integer Phase Ambiguity Resolution:} To resolve the integer ambiguity $m$, the approach exploits the algebraic properties of polar codes under phase rotation. Consider a polar codeword $\mathbf{c} = [c_0, c_1, c_2, c_3, \ldots, c_{N-1}]$. Phase rotations by multiples of $\pi/2$ transform the codeword according to specific patterns:
\begin{align}
\mathbf{c}^{0} &= [c_0, c_1, c_2, c_3, \ldots, c_{N-2}, c_{N-1}], \\
\mathbf{c}^{\pi/2} &= [\bar{c}_1, c_0, \bar{c}_3, c_2, \ldots, \bar{c}_{N-1}, c_{N-2}], \\
\mathbf{c}^{\pi} &= [\bar{c}_0, \bar{c}_1, \bar{c}_2, \bar{c}_3, \ldots, \bar{c}_{N-2}, \bar{c}_{N-1}], \\
\mathbf{c}^{3\pi/2} &= [c_1, \bar{c}_0, c_3, \bar{c}_2, \ldots, c_{N-1}, \bar{c}_{N-2}],
\end{align}
where $\bar{c} = c \oplus 1$ (binary addition) and $\mathbf{c}^{\phi}$ denotes the codeword obtained after phase rotation by $\phi$.

These transformations can be decomposed into a permutation followed by a translation. The permutation $\sigma$ is defined as
\begin{align}
\sigma: [c_0, c_1, c_2, c_3, \ldots] \mapsto [c_1, c_0, c_3, c_2, \ldots], \label{eq:permutation}
\end{align}
which swaps adjacent pairs of bits. The rotated codewords can then be expressed as
\begin{align}
\mathbf{c}^{\pi/2} &= \sigma(\mathbf{c}^{0}) \oplus [1, 0, 1, 0, \ldots, 1, 0], \\
\mathbf{c}^{\pi} &= \sigma(\mathbf{c}^{0}) \oplus [1, 1, 1, 1, \ldots, 1, 1], \\
\mathbf{c}^{3\pi/2} &= \sigma(\mathbf{c}^{0}) \oplus [0, 1, 0, 1, \ldots, 0, 1],
\end{align}
where $\oplus$ denotes bit-wise XOR operation.

According to \cite[Definition 6]{Dragoi-partial-order}, the permutation $\sigma$ corresponds to an affine transformation of the monomial representation that preserves the polar code structure, making it an automorphism for polar codes following the standard partial order.\footnote{The permutation $\sigma$ corresponds to affine transform ${\bf p} \mapsto {\bf A}{\bf p} + {\bf b}$ of vector of monomials ${\bf p} = [p_0, p_1, \ldots, p_{\log_2N-1}]$ with ${\bf A} = {\bf I}$ and ${\bf b} = [0, 0, \ldots, 0, 1]^\top$. The translation corresponds to flipping some message bits as per \eqref{eq:frozen-offset-1}-\eqref{eq:frozen-offset-3}.}

By our assumption that $\{N-2, N-1\} \subseteq \mathcal{F}$, these positions are normally set to zero in the original codeword $\mathbf{c}^0$. However, under phase rotation, these positions take on specific values that uniquely identify the rotation angle:
\begin{align}
\mathbf{c}^{\pi/2} &: [u_{N-2}, u_{N-1}] = [1, 0], \label{eq:frozen-offset-1}\\
\mathbf{c}^{\pi} &: [u_{N-2}, u_{N-1}] = [0, 1], \label{eq:frozen-offset-2}\\
\mathbf{c}^{3\pi/2} &: [u_{N-2}, u_{N-1}] = [1, 1]. \label{eq:frozen-offset-3}
\end{align}

These relationships enable integer phase ambiguity resolution by treating the last two frozen positions as information bits during decoding, then using the decoded values to determine the rotation angle.



\subsection{Coherent Decoding for ${\bf x}_0$}
 The blind decoding architecture enables the receiver to estimate the channel without explicit pilots by treating the decoded codeword $\hat{\mathbf{c}}_1$ as an effective sequence of pilot symbols. Let $\hat{\mathbf{x}}_1$ denote the corresponding QPSK-modulated symbol vector. The receiver computes the maximum likelihood channel estimate as
\begin{align}
\hat{h} = \frac{1}{N_c^1}\sum_{i=0}^{N_c^1-1} y_{1,i}\hat{x}_{1,i}^*, \label{eq:ml_channel_estimate}
\end{align}
with estimation variance $\sigma_{\hat{h}}^2 = \sigma^2/N_c^0$.

This channel estimate is then used for soft demodulation of the remaining data symbols. Specifically, for each bit $k$ of symbol $i$, the receiver computes the log-likelihood ratio (LLR) as
\begin{align}
\text{LLR}_{i,k} = \log \frac{\sum_{x\in\mathcal{K}_{k,1}} \exp\left(-\frac{|y_{0,i} - \hat{h}x|^2}{\sigma_{\text{eff}}^2}\right)}{\sum_{x\in\mathcal{K}_{k,0}} \exp\left(-\frac{|y_{0,i} - \hat{h}x|^2}{\sigma_{\text{eff}}^2}\right)}, \label{eq:llr_computation}
\end{align}
where $\mathcal{K}_{k,0}$ and $\mathcal{K}_{k,1}$ denote the sets of constellation points with the $k$-th bit equal to 0 and 1, respectively, and $\sigma_{\text{eff}}^2 = \sigma_{\hat{h}}^2 + \sigma^2$ captures the aggregate uncertainty from both channel estimation and thermal noise. Armed with these bit-wise LLRs, the receiver proceeds with polar decoding to recover the information bits in $\mathbf{m}_0$.


\subsection{Complexity Analysis}
We analyze the computational complexity of the proposed pilot-free scheme and compare it with conventional pilot-aided transmission. The complexity is evaluated in terms of the number of arithmetic operations required for encoding and decoding.

\subsubsection{Encoding Complexity}

The encoding complexity of the proposed scheme is identical to that of conventional polar codes. Each component code $\mathcal{C}_i$ requires $\mathcal{O}(N_i \log_2 N_i)$ operations for polar encoding, where $N_i$ is the mother code length. Since encoding operations are performed independently for each component, the total encoding complexity is
\begin{equation}
\mathcal{O}(N_0 \log_2 N_0 + N_1 \log_2 N_1).
\end{equation}
This complexity is comparable to conventional pilot-aided schemes using polar codes with similar total codeword lengths.

\subsubsection{Decoding Complexity}

The decoding complexity consists of three main components: blind channel estimation for the QPSK segment, blind decoding of $\mathcal{C}_1$, and coherent decoding of $\mathcal{C}_0$.

\textbf{Channel Magnitude Estimation:} Computing the channel magnitude estimate in~\eqref{eq:magnitude_estimate} requires $N_c^1$ additions for the summation, one square root operation, and several scalar multiplications. The complexity is therefore $\mathcal{O}(N_c^1)$.

\textbf{Fractional Phase Offset Estimation:} The VVPE algorithm in~\eqref{eq:vvpe_estimate} involves computing the fourth power of each received symbol, followed by summation and arctangent operations. This requires $\mathcal{O}(N_c^1)$ multiplications and additions, plus one arctangent computation. The overall complexity remains $\mathcal{O}(N_c^1)$.

\textbf{Blind Decoding of $\mathcal{C}_1$:} Resolving the integer phase ambiguity requires one full polar decoding pass of component code $\mathcal{C}_1$. This introduces an additional $\mathcal{O}(N_1 \log_2 N_1)$ operations. Following successful decoding, we must recover the original message by removing the translation vector and inverting the permutation $\sigma$. The translation removal requires $N_1$ binary XOR operations at the frozen positions $\{N_1-2, N_1-1\}$, while the permutation inversion is accomplished through the polar transform, requiring $\mathcal{O}(N_1 \log_2 N_1)$ operations. Since the decoding complexity dominates these post-processing steps, the total complexity for blind decoding of $\mathcal{C}_1$ is $\mathcal{O}(N_1 \log_2 N_1)$.

\textbf{Channel Re-encoding and Estimation:} After decoding $\mathcal{C}_1$, the decoded message $\hat{\mathbf{m}}_1$ is re-encoded to generate the pilot sequence $\hat{\mathbf{x}}_1$. This re-encoding step requires $\mathcal{O}(N_1 \log_2 N_1)$ operations. The maximum likelihood channel estimate in~\label{eq:ml_channel_estimate} is then computed via correlation, requiring $N_c^1$ complex multiplications and additions. The overall complexity for this stage is $\mathcal{O}(N_1 \log_2 N_1 + N_c^1)$.

\textbf{Coherent Decoding of $\mathcal{C}_0$:} Once the channel estimate $\hat{h}$ is obtained, component code $\mathcal{C}_0$ is decoded using standard SC or SCL decoding. The LLR computation in~\eqref{eq:llr_computation} for higher-order QAM requires evaluating exponentials over all constellation points for each received symbol. For $N_s$-QAM, each symbol requires $\mathcal{O}(N_s)$ operations, resulting in $\mathcal{O}(N_c^0 N_s)$ complexity for demapping. The subsequent polar decoding requires $\mathcal{O}(N_0 \log_2 N_0)$ operations. For practical modulation orders ($N_s \leq 64$), the decoding complexity dominates, yielding an overall complexity of $\mathcal{O}(N_0 \log_2 N_0)$ for this stage.

\subsubsection{Overall Complexity}

Combining all stages, the total decoding complexity of the proposed scheme is
\begin{equation}
\mathcal{O}(N_0 \log_2 N_0 + N_1 \log_2 N_1).
\end{equation}
Compared to pilot-aided transmission using a single polar code of length $N_{\text{PAT}}$ with complexity $\mathcal{O}(N_{\text{PAT}} \log_2 N_{\text{PAT}})$, the proposed scheme introduces an additional decoding pass for the QPSK segment. However, since $N_1 \ll N_0$ in typical configurations optimized for spectral efficiency, the additional complexity overhead is modest. Furthermore, the elimination of explicit pilots allows for smaller total codeword lengths while achieving equivalent or superior performance, potentially offsetting the complexity increase.

For SCL decoding with list size $L$, the complexity scales linearly with $L$, yielding
\begin{equation}
\mathcal{O}(L(N_0 \log_2 N_0 + N_1 \log_2 N_1)).
\end{equation}
In practical URLLC scenarios where small list sizes ($L \leq 8$) suffice for the target reliability, this complexity remains manageable and compatible with low-latency requirements.

\section{Code Optimization}\label{sec:optimization}

\begin{figure}
\centering
    \includegraphics[width=1\linewidth]{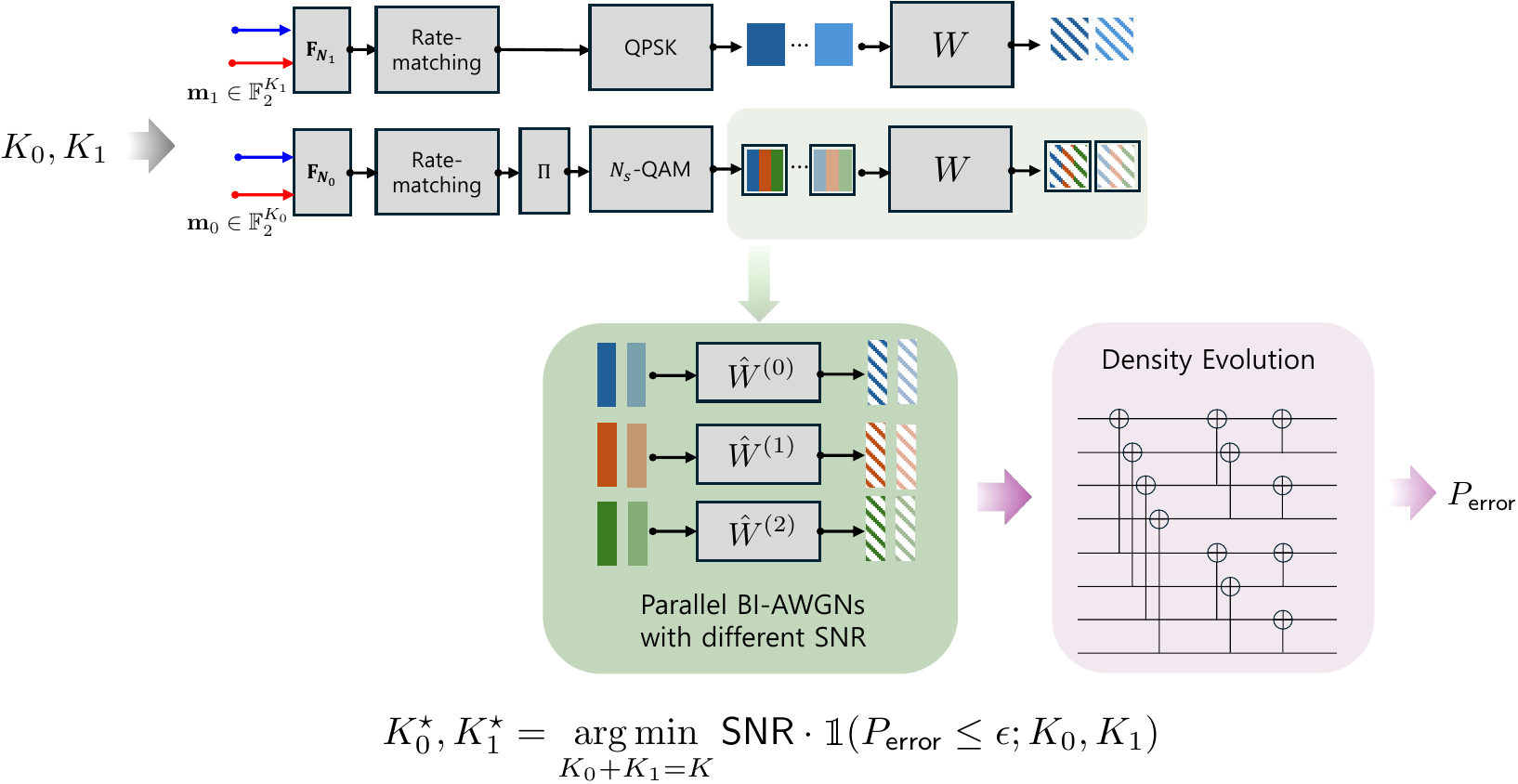}
    \caption{Code parameter optimization procedure. 
    }
    \label{fig:optimization_overall}
\end{figure}

\subsection{BICM channel model}
Consider the complex-AWGN channel $W: \mathcal{X} \rightarrow \mathbb{C}$ with conditional probability density function
\begin{equation}
    W(y | x) = \frac{1}{\pi \sigma^2}e^{-\frac{|y-x|^2}{\sigma^2}}    
\end{equation}
where $\mathcal{X}$ is the constellation symbol set with power normalization $\sum_{x \in \mathcal{X}} |x|^2 = |\mathcal{X}|$, and the signal-to-noise ratio is given by ${\sf SNR} = 1 / \sigma^2$. 

Consider a constellation of size $|\mathcal{X}| = 2^{n_s}$. For each symbol $x_i \in \mathcal{X}$, the corresponding binary sequence ${\bf b}_i = (b_{i,0}, b_{i,1}, \ldots, b_{i,n_s-1}) \in \mathbb{F}_2^{n_s}$ is determined by the bit-labeling function $\psi : \mathbb{F}_2^{n_s} \rightarrow \mathcal{X}$ such that
\begin{align}
    \psi: {\bf b}_i \mapsto x_{{\sf bi2de}({\bf b}_i)},
\end{align}
where ${\sf bi2de}({\bf b}_i) = \sum_{j=0}^{n_s-1} b_{i,j} 2^{j}$ converts the binary vector to its decimal equivalent.

Instead of operating on symbols $x_i$ directly, we define the equivalent channel $\bar{W}$ as $\bar{W}(y | {\bf b}_i) = W(y | \psi({\bf b}_i))$. This channel can be decomposed into $n_s$ binary input channels $\bar{W}^{(k)}: \mathbb{F}_2 \rightarrow \mathbb{C}$, with probability density function
\begin{align}
    &\bar{W}^{(k)}(y | b_{i, k}; b_{i, k-1}, \ldots, b_{i, 0}) \\&= \frac{1}{Z}\sum_{{\bf b} \in \mathbb{F}_2^{n_s - i - 1}}W(y | \psi([{\bf b}, b_{i,k}, b_{i,k-1}, \ldots, b_{i,0}]))
\end{align}
where $b_{i, k-1}, \ldots, b_{i,0}$ represent genie-aided bits corresponding $\phi({\bf b}_i)$ and $Z$ is the normalization factor.

In the BICM channel model, the channel $\bar{W}^{(k)}$ is approximated by $\tilde{W}^{(k)}: \mathbb{F}_2 \rightarrow \mathbb{C}$ with density function
\begin{align}
    &\tilde{W}^{(k)} (y | b_{i,k}) \\ &= \frac{1}{Z}\sum_{{\bf b}_b \in \mathbb{F}_2^{n_s - i - 1}} \sum_{{\bf b}_a \in \mathbb{F}_2^{i}} W(y | \psi([{\bf b}_b, b_{i,k}, {\bf b}_a])).
\end{align}

In terms of mutual information, the BICM channel induces some loss in achievable rate:
\begin{align}
    I(W) &= I(X;Y) = I(B_0, \ldots, B_{n_s-1}; Y) \\
    &= \sum_{k=0}^{n_s-1} I(B_k;Y | B_0, \ldots, B_{k-1}) \\
    &= \sum_{k=0}^{n_s-1} I(\bar{W}^{(k)}) \\
    &\geq \sum_{k=0}^{n_s-1} I(B_k;Y) = \sum_{k=0}^{n_s-1} I(\tilde{W}^{(k)})
\end{align}
where $X$, $Y$, and $B_k$ are random variables corresponding to $x$, $y$, and $b_i$,  $I(W)$ is symmetric mutual information of channel $W$ and $I(X;Y)$ is mutual information of random variables $X$ and $Y$.
According to \cite{BICM-Carie-98}, this approximation results in minimal loss of channel capacity when carefully designed bit-to-symbol mappings are employed, such as Gray mapping. Since we use a demapper that treats bits embedded in the same symbol as independent, the BICM channel model is well-suited for our error probability analysis.


\subsection{BI-AWGN Approximation}\label{subsec:BI-AWGN-approx}
The parallel channel $\tilde{W}^{(k)}$ is non-Gaussian channel in general.  
In our code design, we approximate the channel $\hat{W}^{(k)}$ as binary input AWGN channel with conditional probability density function
\begin{align}
    \hat{W}^{(k)} (y | b) = \frac{1}{\sqrt{2\pi\sigma_k^2}} e^{-\frac{(y - (1-2b))^2}{2\sigma_k^2}},
\end{align}
where $\sigma_k^2$ is corresponding Gaussian noise variance and equivalent SNR is $1/\sigma_k^2$.

To compute the equivalent SNR, we match the mutual information of the two channels as
\begin{align}
   I(\hat{W}^{(k)}) = I(\tilde{W}^{(k)}).
\end{align}
Intuitively, this matching process preserves the average quality of LLR values for each channel output. In our equivalent BI-AWGN framework, the original symbol-input channel is transformed into parallel bit-input channels with Gaussian-approximated outputs, where the mutual information matching criterion ensures that the average information content available for decoding is maintained despite the non-Gaussian to Gaussian approximation. This approach enables polar code construction and performance analysis. Fig.~\ref{fig:optimization_overall} summarizes the equivalent channel model.


\subsection{Error Probability Analysis}\label{subsec:theory}

Suppose the block error probability of the $i$th component code $\mathcal{C}_i$ is $P_{\sf error}^{(i)}$. An error occurs if at least one component code fails. Thus, the overall error probability $P_{\sf error}$ is computed as
\begin{align}
    P_{\sf error} = 1 - (1- P_{\sf error}^{(0)})(1- P_{\sf error}^{(1)}).
\end{align}

To compute the error probability of each component code, we adopt the density evolution framework with Gaussian approximation, abbreviated as \textit{DEGA}\cite{Trifnov-polar-construction, Wu-DEGA-SC-Pe}. Consider a polar code with mother code length $N$. For simplicity, we ignore the interleaver in the following explanation.

The $i$th codeword bit $c_i = c_{n_s k + j}$ is modulated into the $k$th transmitted symbol and transmitted over the $j$th equivalent BI-AWGN channel $\hat{W}^{(j)}$. The channel log-likelihood ratio (LLR) for the $i$th codeword bit is defined as 
\begin{equation}
    L_{i}^{(c)} = \log\frac{P(y_k | c_i=0)}{P(y_k|c_i=1)}.
\end{equation}

The successive cancellation (SC) decoder uses the channel LLRs $(L_0^{(c)}, L_1^{(c)}, \ldots, L_{N-1}^{(c)})$ to compute the bit LLRs $(L_0^{(u)}, L_1^{(u)}, \ldots, L_{N-1}^{(u)})$, defined as
\begin{equation}
    L_i^{(u)} = \log \frac{ P({\bf y}, {\bf u}_0^{i-1} | u_i = 0) }{ P({\bf y}, {\bf u}_0^{i-1} | u_i = 1) }.
\end{equation}

In our error analysis, we focus on the propagation from channel LLRs to bit LLRs, which is directly used to compute the code block error probability. We begin by reviewing the LLR propagation process during SC decoding.

\subsubsection{LLR Propagation Process in SC Decoding}
SC decoding can be understood as a message passing algorithm that follows a specific scheduling policy for processing elements.

{\bf Processing Elements:}
To explain the decoding process, we introduce the processing element, which is a fundamental component of the polar transform. Each processing element takes a pair of scalar inputs $(U_1, U_2)$ and produces a pair of scalar outputs $(U_1 \oplus U_2, U_2)$, where $\oplus$ denotes the XOR operation. The polar transform consists of $\log_2(N)$ levels, and each level contains $N/2$ processing elements arranged in parallel.

{\bf Labeling and Message Flow:}
We assign the label $(d, j)$ to the $j$th processing element at the $d$th level, where $d \in \{0, 1, \ldots, \log_2(N)-1\}$ and $j \in \{0, 1, \ldots, N/2-1\}$. Each processing element handles two types of messages with $k \in \{1,2\}$ indicating the upper or lower position:

\begin{itemize}
\item \textbf{Hard-decision bits:} $U_{j,k}^{d,\rightarrow}$ (inputs from left) and $U_{j,k}^{d}$ (outputs to right)
\item \textbf{LLR messages:} $L_{j,k}^{d,\leftarrow}$ (inputs from right) and $L_{j,k}^{d}$ (outputs to left)
\end{itemize}
Hard-decision bits and LLR messages flow in opposite directions through the polar decoding tree. Hard-decision bits propagate from the decoded information bits (leftmost level) back toward the channel (rightmost level), hence the $\rightarrow$ notation for inputs. Conversely, LLR values propagate from the channel outputs (rightmost level) toward the information bits (leftmost level), hence the $\leftarrow$ notation for LLR inputs. 

To establish the correspondence between processing elements and codeword bits, we introduce a mapping function $\alpha_d: (j,k) \mapsto i$ that maps processing element position $(j,k)$ at level $d$ to codeword bit index $i$: 
\begin{align}
    \alpha_d(j,k) &= \left\lfloor\frac{j}{2^{n-d-1}}\right\rfloor2^{n-d}  + \left( j - \left\lfloor\frac{j}{2^{n-d-1}}\right\rfloor 2^{n-d-1}\right) \nonumber  \\
    & \quad\quad+(k-1)2^{n-d-1}
\end{align}






The systematic interconnection between processing elements at different levels is defined by the permutation relationships:
\begin{align}
    U_{\alpha_{d}^{-1}(\alpha_{d-1}(j,k))}^{d,\rightarrow} &= U_{j,k}^{d-1} \\
    L_{\alpha_{d+1}^{-1}(\alpha_d(j,k))}^{d+1,\leftarrow} &= L_{j,k}^d,
\end{align}
where $\alpha_d^{-1}$ denotes the inverse mapping that converts codeword bit index back to processing element position at level $d$.

{\bf Message Updates:}
Each processing unit uses LLR messages from the right (higher-indexed levels) to compute LLR outputs toward the left (lower-indexed levels):
\begin{align}
    L_{\alpha_{d+1}^{-1}(\alpha_d(j,1))}^{d+1,\leftarrow} = L_{j,1}^d &= f(L_{j,1}^{d,\leftarrow}, L_{j,2}^{d,\leftarrow}), \label{eqn:l2r-first}\\
    L_{\alpha_{d+1}^{-1}(\alpha_d(j,2))}^{d+1,\leftarrow} = L_{j,2}^d &= g(L_{j,1}^{d,\leftarrow}, L_{j,2}^{d,\leftarrow}, U_{j,1}^{d,\rightarrow}), \label{eqn:l2r-second}
\end{align}
where the functions $f$ and $g$ are defined as
\begin{align}
    f(L_a, L_b) &= \log\left(\frac{1+e^{L_a+L_b}}{e^{L_a} + e^{L_b}} \right), \\
    g(L_a, L_b, u) &= L_b + (1-2u) L_a.
\end{align}
Similarly, each processing unit propagates hard-decision bits from the left (lower-indexed levels) toward the right (higher-indexed levels):
\begin{align}
    U_{\alpha_{d-1}^{-1}(\alpha_{d}(j,1))}^{d-1,\rightarrow} = U_{j,1}^{d} &= U_{j,1}^{d,\rightarrow} \oplus U_{j,2}^{d,\rightarrow}, \\
    U_{\alpha_{d-1}^{-1}(\alpha_{d}(j,2))}^{d-1,\rightarrow} = U_{j,2}^{d} &= U_{j,2}^{d,\rightarrow}.
\end{align}

The message updates are performed only when all required messages are available. The hard-decision bits $U_{j,k}^{n,\rightarrow}$ are initialized as
\begin{equation}
    U_{j,k}^{n,\rightarrow} = \begin{cases}
        0, & \alpha_n(j,k)\in\mathcal{F} \\
        0, & \alpha_n(j,k) \in \mathcal{I} \text{ and } L_{j,k}^{n} > 0 \\
        1, & \alpha_n(j,k) \in \mathcal{I} \text{ and } L_{j,k}^{n} < 0.
    \end{cases}
\end{equation}
The LLR values are initialized as $L_{j,k}^{0,\leftarrow} = L_{\alpha_0(j,k)}^{(c)}$.

\subsubsection{Statistical Characterization}

Instead of computing exact LLR values, DEGA models the LLR messages as Gaussian random variables and tracks their means $\mu_{j,k}^{d}$ and variances $(\sigma_{j,k}^{d})^2$:
\begin{equation}
    L_{j,k}^{d,\leftarrow} \sim \mathcal{N}(\mu_{j,k}^{d}, (\sigma_{j,k}^{d})^2).
\end{equation}

In the BICM channel model with BI-AWGN approximation, we assume that each codeword bit $c_i$ is transmitted through an AWGN channel with different channel qualities. Due to channel symmetry, the variance of the LLR is determined by its mean through the relation $(\sigma_{j,k}^{d})^2 = \mu_{j,k}^{d}$. This approximation greatly simplifies the analysis while providing accurate estimates of the decoding error probability under SC decoding.

DEGA is equivalent to performing SC decoding under the assumption that all bits are frozen (i.e., $U_{j,k}^{d,\rightarrow}=0$), but with modified message passing rules and initialization procedures. We first introduce the equivalent channel model and present the initialization process based on this model, followed by the modified message passing rules for statistical tracking.

{\bf Codeword Bits to Symbols:}
Consider the codeword ${\bf c} = (c_0, c_1, \ldots, c_{M-1})$ with length $M$ and a constellation of size $2^{n_s}$. Before symbol mapping, the codeword bits are typically interleaved as 
\begin{equation}
    (c_0, c_1, \ldots, c_{M-1}) \mapsto (c_{\pi(0)}, c_{\pi(1)}, \ldots, c_{\pi(M-1)}),
\end{equation}
where $\pi(\cdot)$ is the interleaving permutation.
The interleaved bits are then grouped into $n_s$ distinctive sets according to their bit positions within each symbol:
\begin{equation}
    \mathcal{B}_i = \{c_{\pi(i)}, c_{\pi(n_s+i)}, c_{\pi(2n_s+i)}, \ldots, c_{\pi((B-1)n_s + i)}\},
\end{equation}
where $B = M/n_s$ is the number of transmitted symbols and $i \in \{0, 1, \ldots, n_s-1\}$.
The symbol mapper $\phi:\mathcal{B}_0 \times \mathcal{B}_1 \times \cdots \times \mathcal{B}_{n_s-1} \rightarrow \mathcal{X}$ takes $n_s$ bits from the codeword to form the $j$th transmitted symbol:
\begin{equation}
    \phi: {\bf c}_j \mapsto x_{\sf bi2de({\bf c}_j)},
\end{equation}
where ${\bf c}_j = (c_{\pi(jn_s)}, c_{\pi(jn_s+1)}, \ldots, c_{\pi((j+1)n_s -1)})$ and the symbol is transmitted through the AWGN channel.

{\bf BI-AWGN Channel Model:}
According to the BI-AWGN approximation model introduced in Section~\ref{subsec:BI-AWGN-approx}, the bits in set $\mathcal{B}_i$ are treated as being transmitted through an independent equivalent AWGN channel $\hat{W}^{(i)}$ with noise variance $\sigma_i^2$. The input-output relationship of the $i$th equivalent channel is
\begin{equation}
    \hat{W}^{(i)}: \mathcal{B}_i  \rightarrow \mathbb{R}, \quad
    x \mapsto y,
\end{equation}
with probability density function
\begin{equation}
    \hat{W}^{(i)}(y | x) = \frac{1}{\sqrt{2\pi\sigma_i^2}} e^{-\frac{(y-x)^2}{2\sigma_i^2}}.
\end{equation}

{\bf LLR Initialization:}
Using the equivalent channels $\hat{W}^{(i)}$, we can compute the average LLR values from the channel observations. For the zero-codeword transmission assumption, the average LLR for bits transmitted through channel $\hat{W}^{(i)}$ is $\frac{2}{\sigma_i^2}$.
To establish the correspondence between polar decoding labels and channel outputs, we initialize the LLR means at the rightmost level (level $n = \log_2(N)$). If the codeword bit corresponding to position $(j,k)$ at level $n$ is transmitted through channel $\hat{W}^{(p)}$, we set:
\begin{align}
    \mu_{j,k}^{n} = \frac{2}{\sigma_p^2}, \quad \pi(\alpha_n(j,k)) \in \mathbb{Z} + p.
\end{align}

{\bf Statistical Message Updates:}
The update rules directly follow from the SC decoding message update rules. At each processing element, equation \eqref{eqn:l2r-first} can be rewritten as
\begin{equation}
    \tanh\left(\frac{1}{2}L_{\alpha_{d+1}^{-1}(\alpha_d(j,1))}^{d+1,\leftarrow}\right) = \tanh\left(\frac{1}{2}L_{j,1}^{d,\leftarrow}\right)\tanh\left(\frac{1}{2}L_{j,2}^{d,\leftarrow}\right).
\end{equation}
Under the Gaussian approximation where $L_{j,k}^{d,\leftarrow} \sim \mathcal{N}(\mu, 2\mu)$, we derive the mean update equations by taking expectations. The expectation can be expressed using the function $\psi(\cdot)$, defined as 
\begin{equation}
    \psi(\mu) = \mathbb{E}\left[ \tanh\left( \frac{X}{2} \right) \right],    
\end{equation}
where $X \sim \mathcal{N}(\mu, 2\mu)$. This function can be approximated as \cite{Chung-DEGA-01}
\begin{equation}
\psi(x) \approx \begin{cases} 
1-\exp(-0.4527x^{0.86} + 0.0218), &  0 < x \le 10,\\
1-\sqrt{\frac{\pi}{x}}\left(1 - \frac{10}{7x}\right) e^{-\frac{x}{4}}, &  x > 10.
\end{cases}
\end{equation}

For the second output corresponding to equation \eqref{eqn:l2r-second}, under the assumption that all bits are frozen (i.e., $U_{j,1}^{d,\rightarrow} = 0$), the update simplifies to a direct addition of the input means.

Therefore, at each processing element, the Gaussian distribution parameters are updated according to:
\begin{align}
    \mu_{\alpha_{d+1}^{-1}(\alpha_d(j,1))}^{d+1} &= \psi^{-1}( \psi(\mu_{j,1}^{d}) \cdot \psi(\mu_{j,2}^{d})), \\
    \mu_{\alpha_{d+1}^{-1}(\alpha_d(j,2))}^{d+1} &= \mu_{j,1}^{d} + \mu_{j,2}^{d}, \\
    \sigma_{j,k}^{d} &= \sqrt{2\mu_{j,k}^{d}}.
\end{align}

These update rules enable efficient tracking of LLR statistics throughout the polar decoding tree without requiring actual message computations, significantly reducing the computational complexity of error probability analysis.

\subsubsection{Block Error Probability of SC Decoding}
The SC decoding error probability for $\mathcal{C}_i$ can be approximated following \cite{Trifnov-polar-construction, Wu-DEGA-SC-Pe} with the modified channel initialization as:
\begin{align}
    P_{\sf error}^{(i)} &= 1 - \prod_{j,k: \alpha_n(j,k) \in \mathcal{I}_i} \left( 1 - Q\left( \sqrt{\frac{\mu_{j,k}^{n}}{2}} \right) \right),
    \label{eqn:P-error-v1}
\end{align}
where $\mathcal{I}_i$ denotes the set of information bit indices for $\mathcal{C}_i$, and $\mu_{j,k}^{n}$ represents the final LLR mean at the leftmost level (level $n = \log_2(N)$) corresponding to the information bit decisions. The product is taken over all processing element positions $(j,k)$ at level $n$ whose corresponding codeword bit indices $\alpha_n(j,k)$ belong to the information set $\mathcal{I}_i$.

For the proposed code-splitting scheme, we compute the error probabilities with appropriate channel modifications:
\begin{itemize}
\item For $\mathcal{C}_1$: Standard DEGA initialization at level 0 using equivalent channel variances $\sigma_p^2$
\item For $\mathcal{C}_0$: Modified initialization at level 0 using $\sigma_p^2 + \sigma_{\hat{h}}^2$ to account for channel estimation error variance $\sigma_{\hat{h}}^2$
\end{itemize}

The overall error probability is then computed as:
\begin{align}
    P_{\sf error} = 1 - (1- P_{\sf error}^{(0)})(1- P_{\sf error}^{(1)}).
\end{align}
This approach captures the trade-off between the improved channel estimation quality (due to the additional reference symbols from successfully decoded $\mathcal{C}_1$) and the reduced code rate for $\mathcal{C}_0$, enabling accurate prediction of the overall system performance under the proposed code-splitting scheme for higher-order QAM.

\subsection{Parameter Optimization for Code-Splitting Design}
Given the total number of channel uses $N_c$ and the allocation $(N_c^0, N_c^1)$ where $N_c^0 + N_c^1 = N_c$, we assume that $N_c^1$ symbols use QPSK modulation while $N_c^0$ symbols employ higher-order QAM modulation. We further assume that the rate-matching method and mother code lengths $N_i$ for each component code are predetermined. Under these constraints, we present a systematic approach to determine the optimal information bit allocation $(K_0, K_1)$ for the component codes $\mathcal{C}_0$ and $\mathcal{C}_1$, summarized in Fig.~\ref{fig:optimization_overall}.

The optimization procedure searches over all feasible combinations of $(K_0, K_1)$ that satisfy the constraints $K_i \leq N_i$ and $K_0 + K_1 = K$, where $K$ is the total number of information bits. For each combination, we compute the BLER using the error probability analysis framework presented in the previous section.

The search algorithm proceeds as follows: First, we define a design SNR range $[\text{SNR}_{\min}, \text{SNR}_{\max}]$ and start the search from $\text{SNR}_{\min}$ with sequential SNR increments. For each SNR value, we evaluate the overall BLER for all feasible $(K_0, K_1)$ combinations using the error probability formula $P_{\sf error} = 1 - (1- P_{\sf error}^{(0)})(1- P_{\sf error}^{(1)})$. Among all combinations, we select the $(K_0^*, K_1^*)$ that achieves the minimum BLER. If this minimum BLER falls below the target threshold, we terminate the search and adopt the corresponding parameters for code design. Otherwise, we proceed to the next SNR value. 
The algorithm finds the optimal parameter pair $(K_0^*, K_1^*)$ that achieves the target BLER at the minimum required SNR under SC decoding.

When cyclic redundancy check (CRC) codes with $K_{\sf crc}$ bits are employed for error detection, the total information bits are adjusted to $K + K_{\sf crc}$, and the same optimization procedure applies with the modified constraint $K_0 + K_1 = K + K_{\sf crc}$. Although the error probability analysis is derived for SC decoding, the optimized parameters can be directly applied to SCL decoding, which typically achieves better performance due to its enhanced error correction capability.

\section{Simulation Results}\label{sec:simulation}

We evaluate the proposed coded-pilot scheme through Monte Carlo simulations under block-fading channels. The channel coefficient $h = e^{j\phi}$ has unit magnitude with uniformly distributed phase $\phi \sim \mathcal{U}(0, 2\pi)$. To enable fair comparison across modulation orders, we fix the total number of channel uses: for a given symbol count, 4-QAM transmits over $x$ channel uses, while 16-QAM uses $x/2$ and 64-QAM uses $x/4$, matching typical spectral efficiency configurations in 5G NR channels \cite{3gpp-nr-modulation}.

All schemes transmit $K$ information bits per packet with an 11-bit CRC for error detection following the 5G NR standard \cite{3gpp-nr-coding}. For the pilot-aided baseline, we optimize pilot allocation by evaluating candidate lengths of 4, 8, 16, and 32 symbols per packet. The pilot length minimizing the required SNR to achieve BLER of $10^{-3}$ is selected for each configuration. The proposed coded-pilot scheme uses the same candidate set $\{4, 8, 16, 32\}$ for $N_1$ selection, with code parameters optimized according to the method in Section~\ref{sec:optimization}. Rate-matching via puncturing (for $R < 0.55$) or shortening (for $R \geq 0.55$). Sub-block interleaving follow 5G NR recommendations.

\subsection{Validation of Theoretical Analysis}

\begin{figure*}[t]
    \centering
    \includegraphics[width=0.32\textwidth]{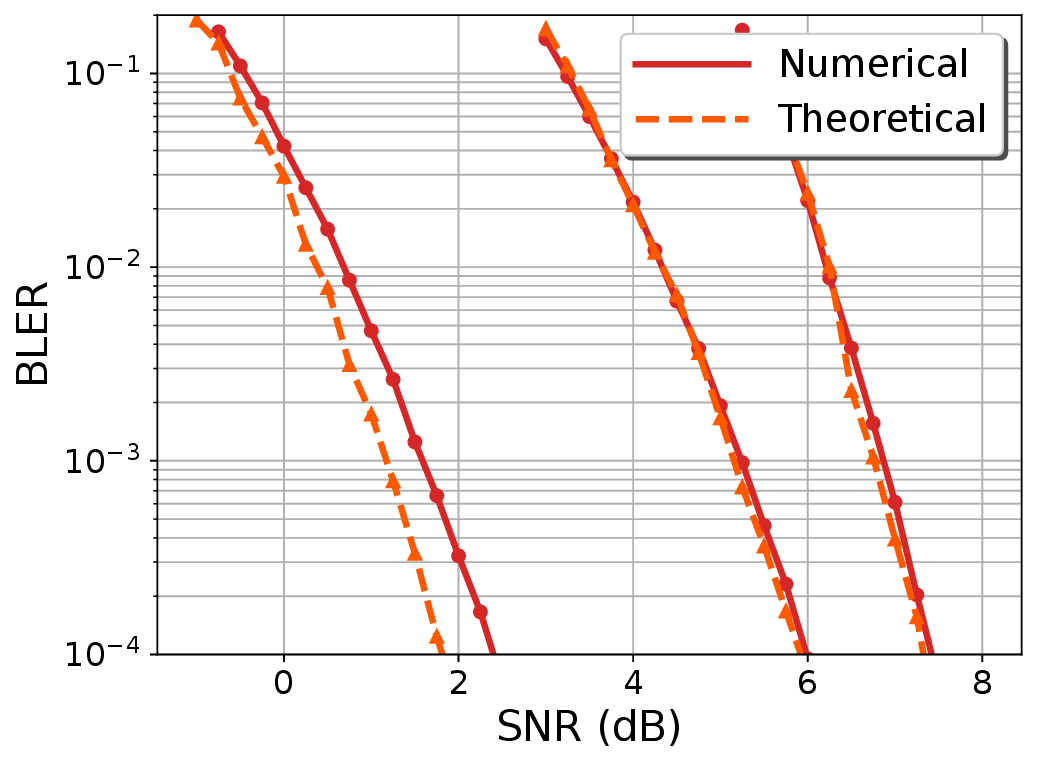}
    \hfil
    \includegraphics[width=0.32\textwidth]{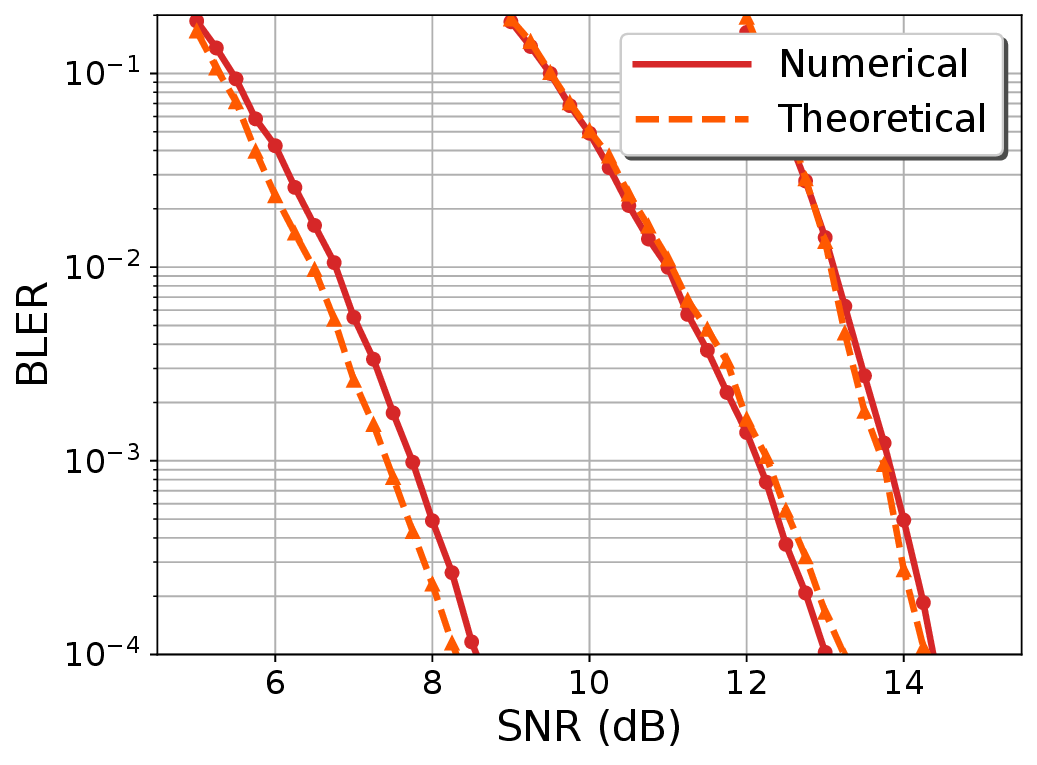}
    \hfil
    \includegraphics[width=0.32\textwidth]{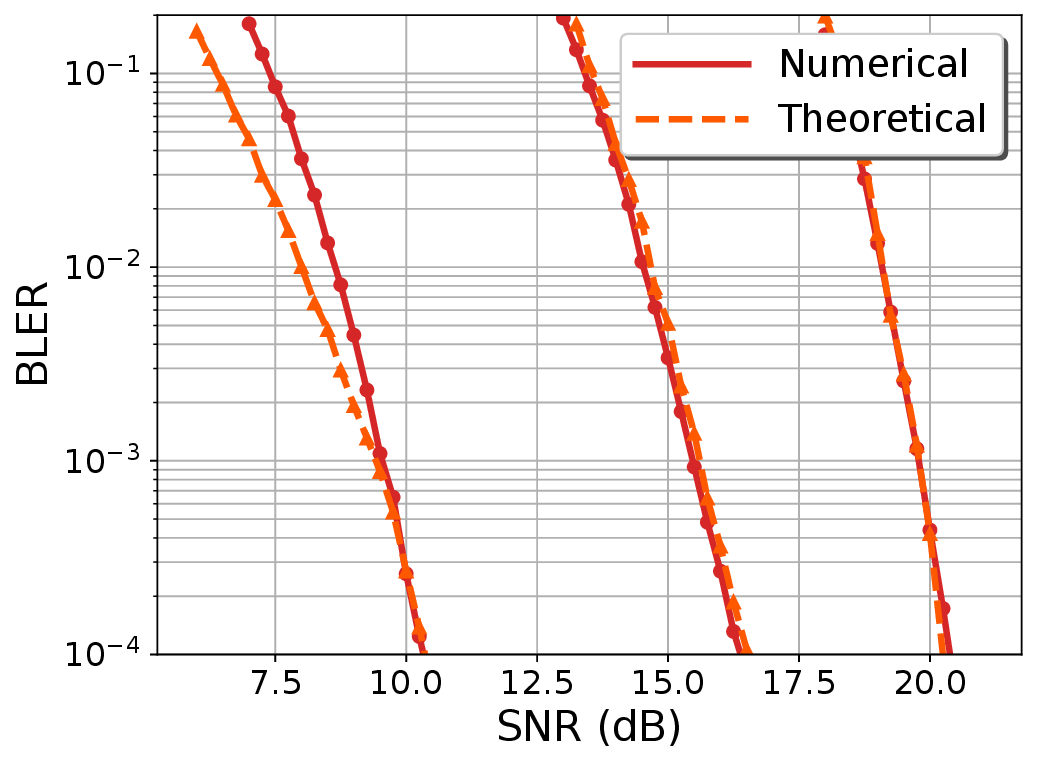}
    
    
    \parbox{0.32\textwidth}{\centering \footnotesize (a) 4-QAM}
    \hfil
    \parbox{0.32\textwidth}{\centering \footnotesize (b) 16-QAM}
    \hfil
    \parbox{0.32\textwidth}{\centering \footnotesize (c) 64-QAM}
    \caption{Comparison between theoretical BLER approximation (dashed lines) and Monte Carlo simulation (solid lines) for single block fading ($B=1$) with $M=600$ total coded bits and $M_1=32$ coded-pilot bits. Three code rates are shown: $R \in \{0.25, 0.5, 0.75\}$ bits/channel use. The theoretical analysis accurately predicts performance across all modulation orders and code rates.}
    \label{fig:BLERTheory_B1}
\end{figure*}
\begin{figure*}[t]
    \centering
    \includegraphics[width=0.32\textwidth]{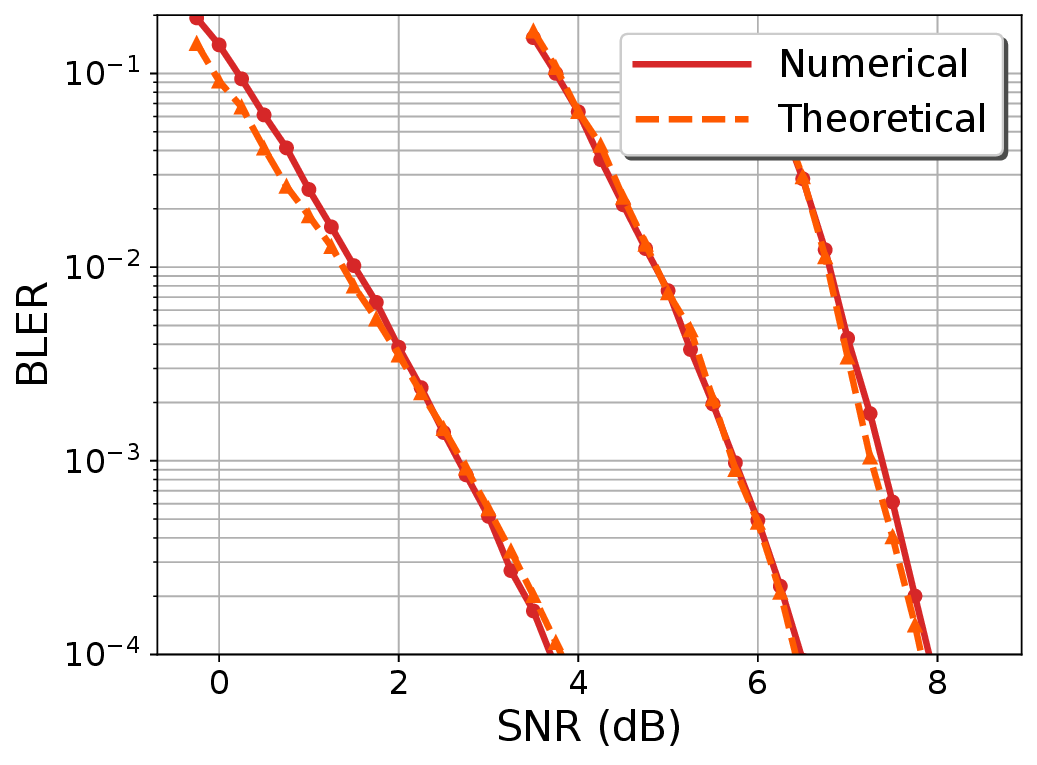}
    \hfil
    \includegraphics[width=0.32\textwidth]{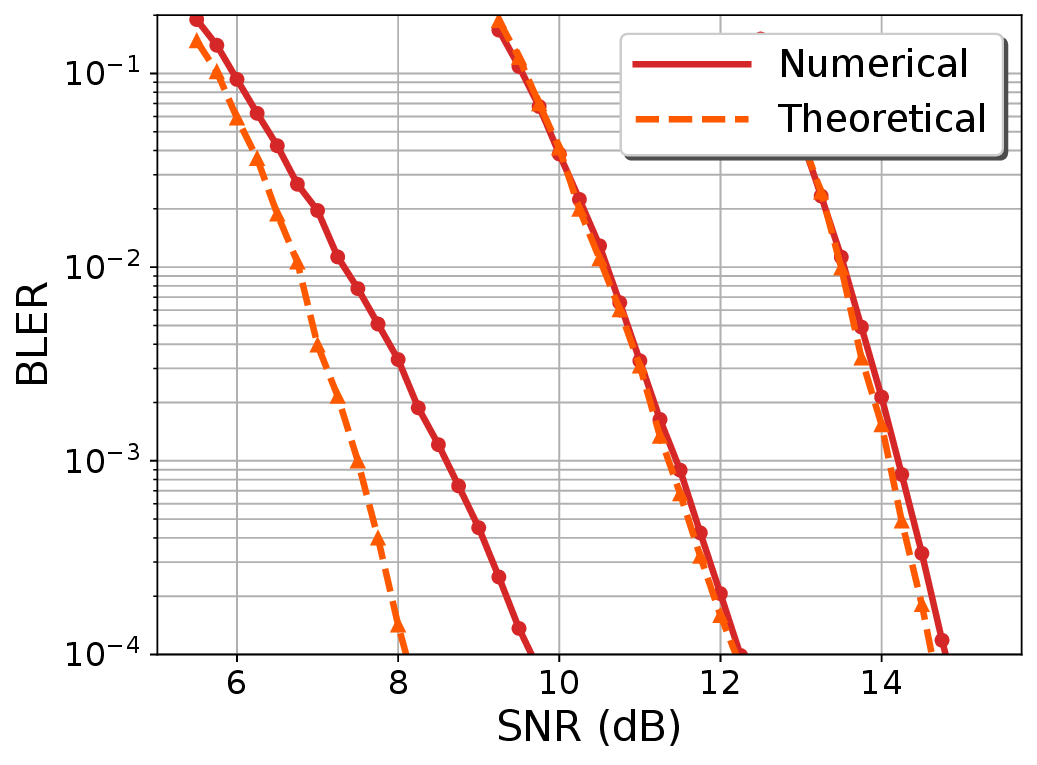}
    \hfil
    \includegraphics[width=0.32\textwidth]{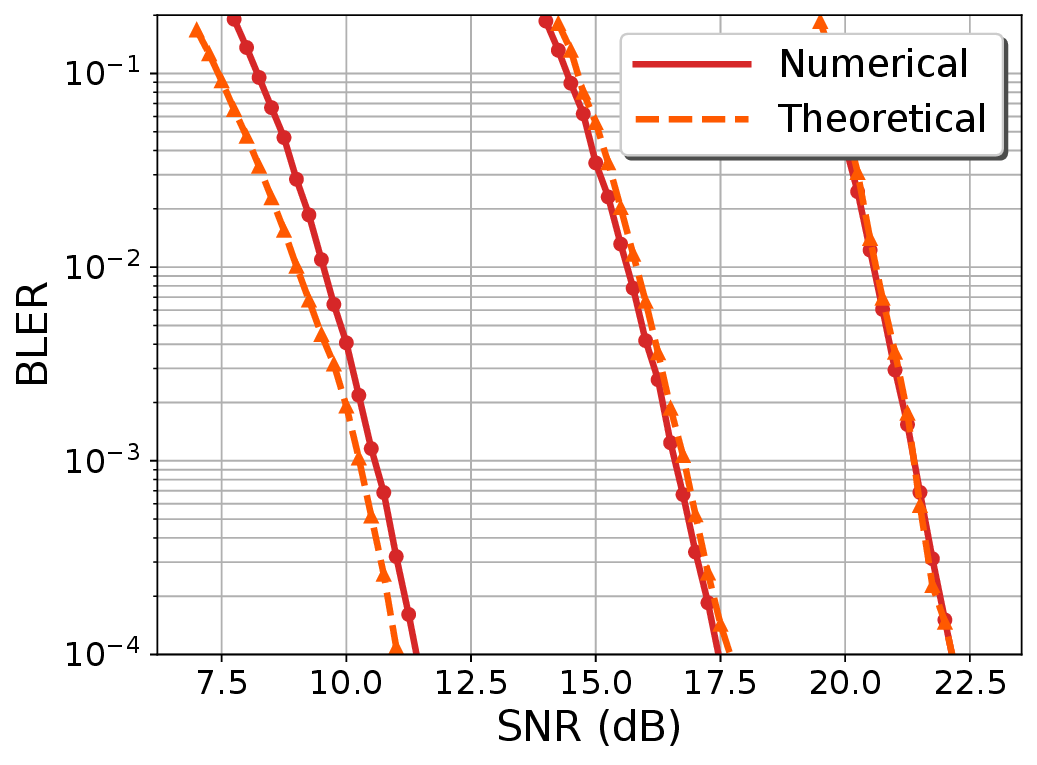}
    
    
    \parbox{0.32\textwidth}{\centering \footnotesize (a) 4-QAM}
    \hfil
    \parbox{0.32\textwidth}{\centering \footnotesize (b) 16-QAM}
    \hfil
    \parbox{0.32\textwidth}{\centering \footnotesize (c) 64-QAM}
    \caption{Comparison between theoretical BLER approximation and Monte Carlo simulation for three block fading ($B=3$) with $M=600$ coded bits and $M_b=16$ coded bits per block. The theoretical model maintains accuracy in multi-block fading scenarios across different modulation orders and code rates.}
    \label{fig:BLERTheory_B3}
\end{figure*}

We first validate the theoretical BLER approximation derived in Section~\ref{subsec:theory} by comparing it against Monte Carlo simulations. SCL decoding with list size $L=1$ is employed. Fig.~\ref{fig:BLERTheory_B1} and Fig.~\ref{fig:BLERTheory_B3} show results for single block fading ($B=1$) and three block fading ($B=3$), respectively, with $M=600$ coded bits and pilot code length $M_1=32$ (equivalently $N_c^1=32$ for 4-QAM, $N_c^1=16$ for 16-QAM, $N_c^1=8$ for 64-QAM).

For each modulation order, we examine three code rates: $R \in \{0.25, 0.5, 0.75\}$ bits per channel use, corresponding to $K \in \{150, 300, 450\}$ information bits. The theoretical curves (dashed lines) closely match the numerical Monte Carlo results (solid lines) across all configurations at medium to high code rates ($R \geq 0.5$). The approximation maintains accuracy over the BLER range from $10^{-4}$ to $10^{-1}$, validating the analytical framework for both single and multiple block-fading scenarios. At the lowest code rate ($R=0.25$), a noticeable gap emerges between theoretical and numerical curves, particularly at low SNR. This discrepancy suggests that the approximation may overestimate performance at low SNR, potentially resulting in suboptimal code parameter selection by the optimization algorithm. Nevertheless, the theoretical model correctly captures the qualitative behavior and provides accuracy acceptable for design purposes, enabling reliable parameter optimization at practical operating points for moderate to high code rates.





\subsection{Performance Comparison with Pilot-Aided Scheme}

\begin{figure*}[t]
    \centering
    
    \includegraphics[width=0.32\textwidth]{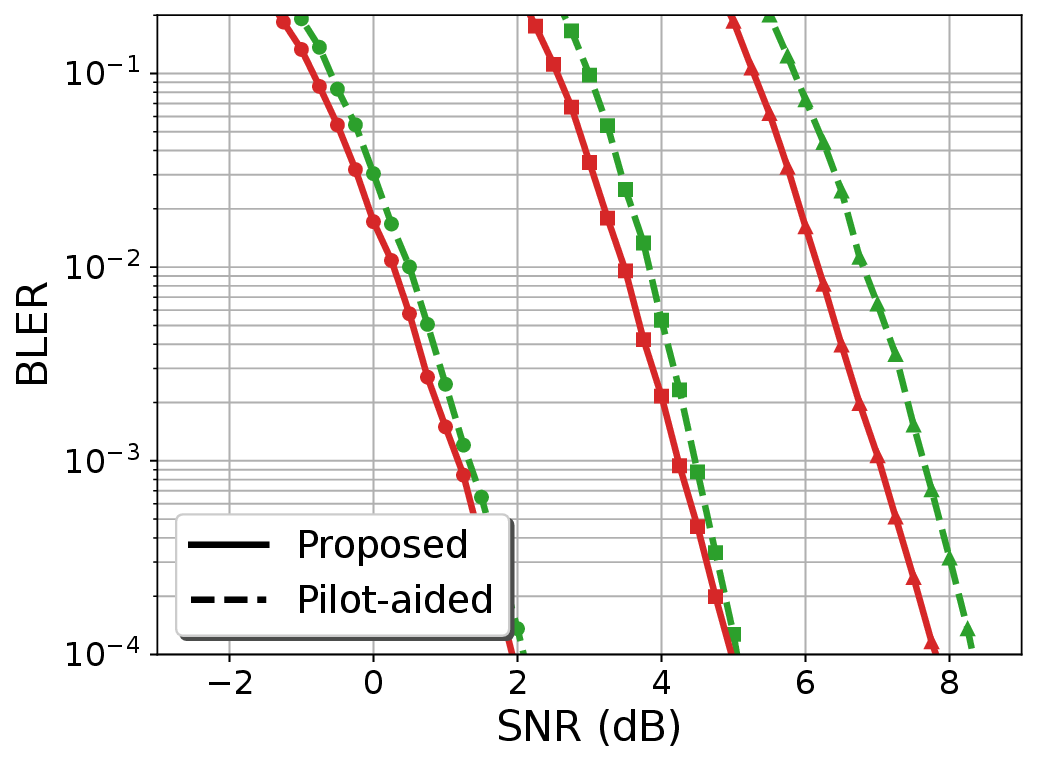}
    \hfil
    \includegraphics[width=0.32\textwidth]{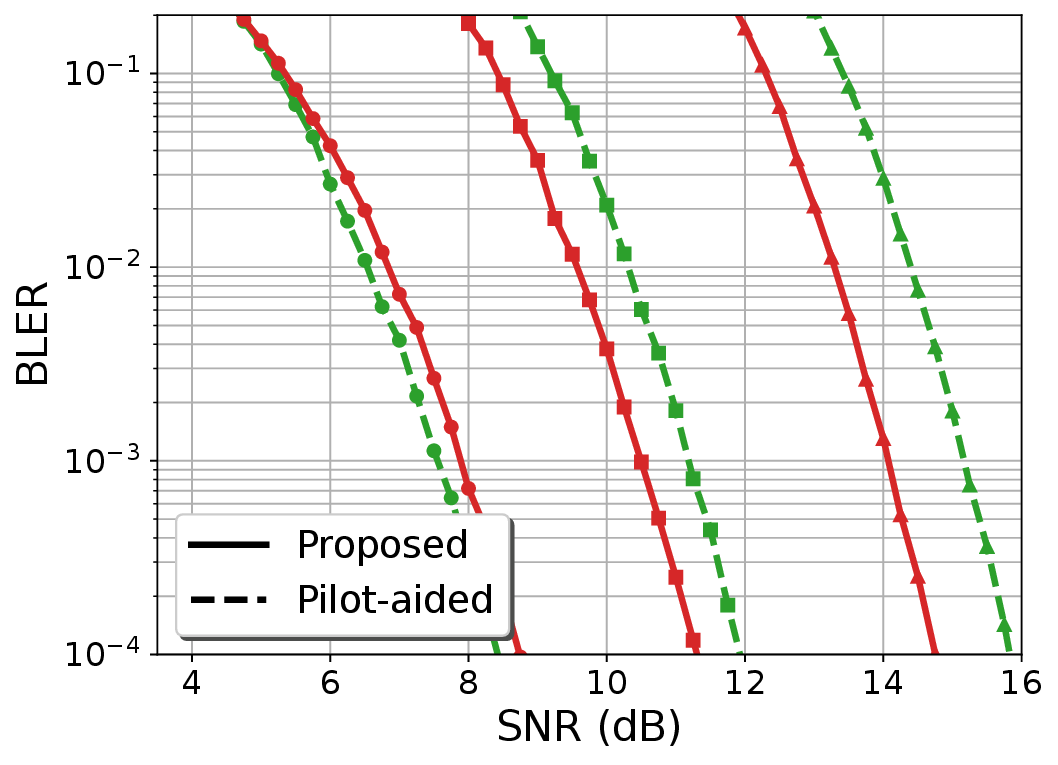}
    \hfil
    \includegraphics[width=0.32\textwidth]{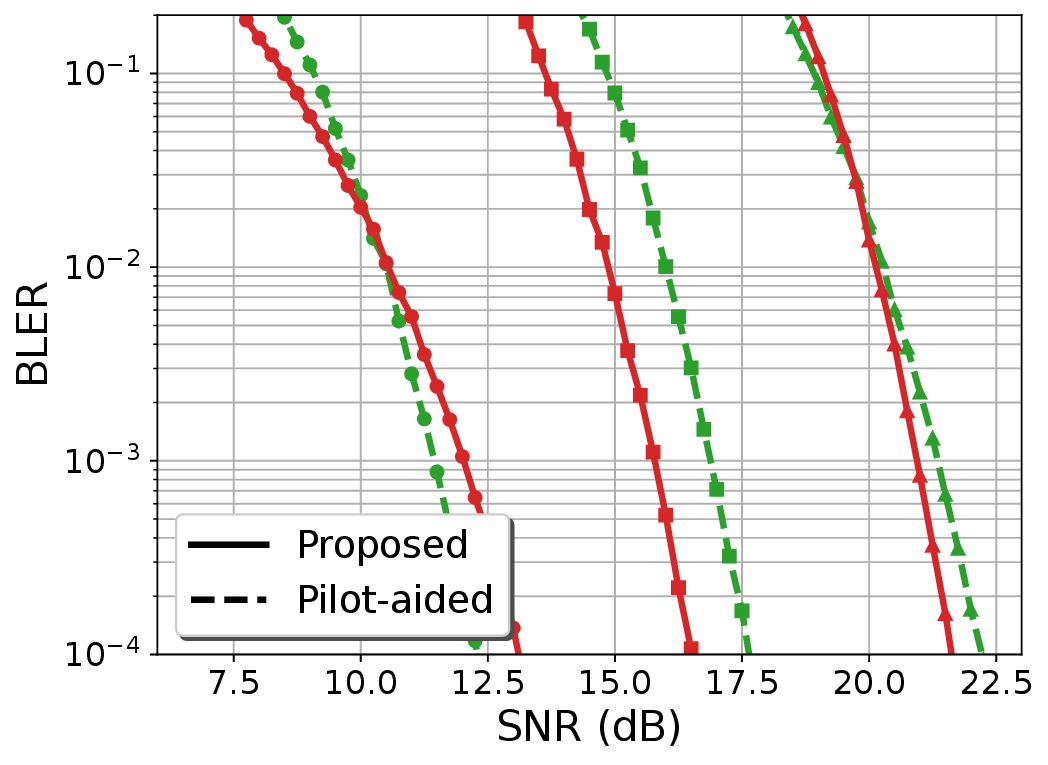}

    \parbox{0.32\textwidth}{\centering \footnotesize (a) 4-QAM}
    \hfil
    \parbox{0.32\textwidth}{\centering \footnotesize (b) 16-QAM}
    \hfil
    \parbox{0.32\textwidth}{\centering \footnotesize (c) 64-QAM}
    
    \caption{BLER performance comparison between proposed coded-pilot scheme (solid lines) and pilot-aided baseline (dashed lines) for single block fading ($B=1$) with $M=240$ coded bits. Three message lengths $K \in \{60, 120, 180\}$ corresponding to code rates $\{0.25, 0.5, 0.75\}$ are evaluated. The proposed scheme achieves consistent SNR gains across all modulation orders, with larger gains at higher code rates.}
    \label{fig:experiments_BLER_B1}
\end{figure*}
\begin{figure*}[t]
    \centering
    
    \includegraphics[width=0.32\textwidth]{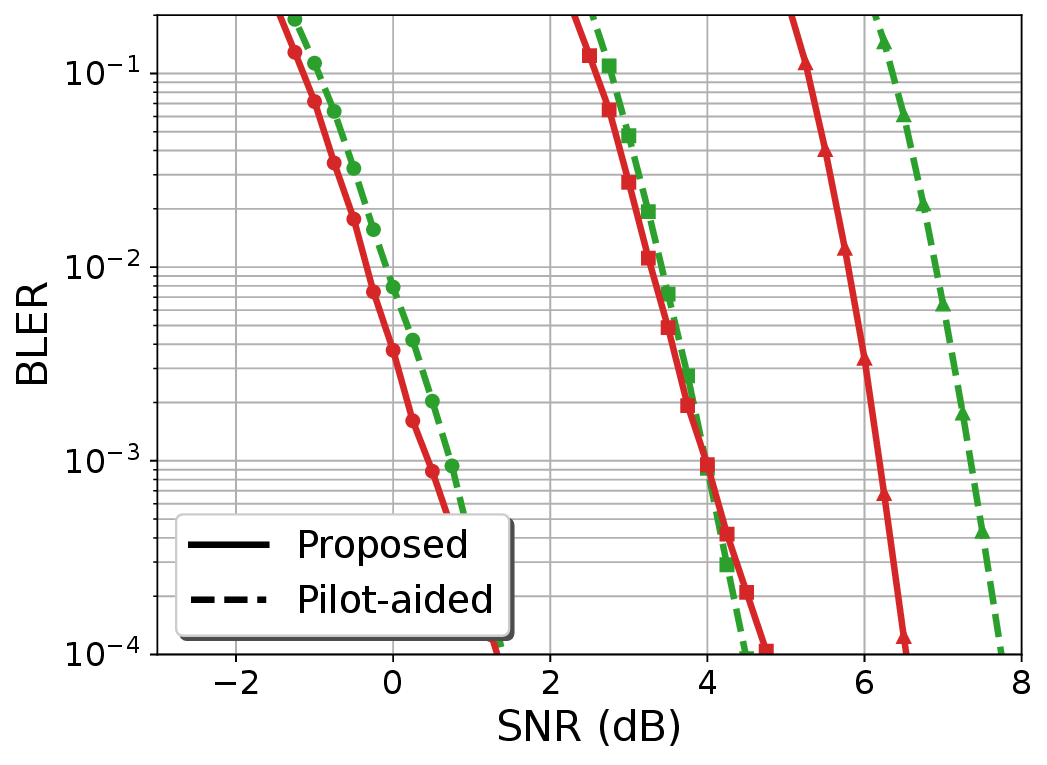}
    \hfil
    \includegraphics[width=0.32\textwidth]{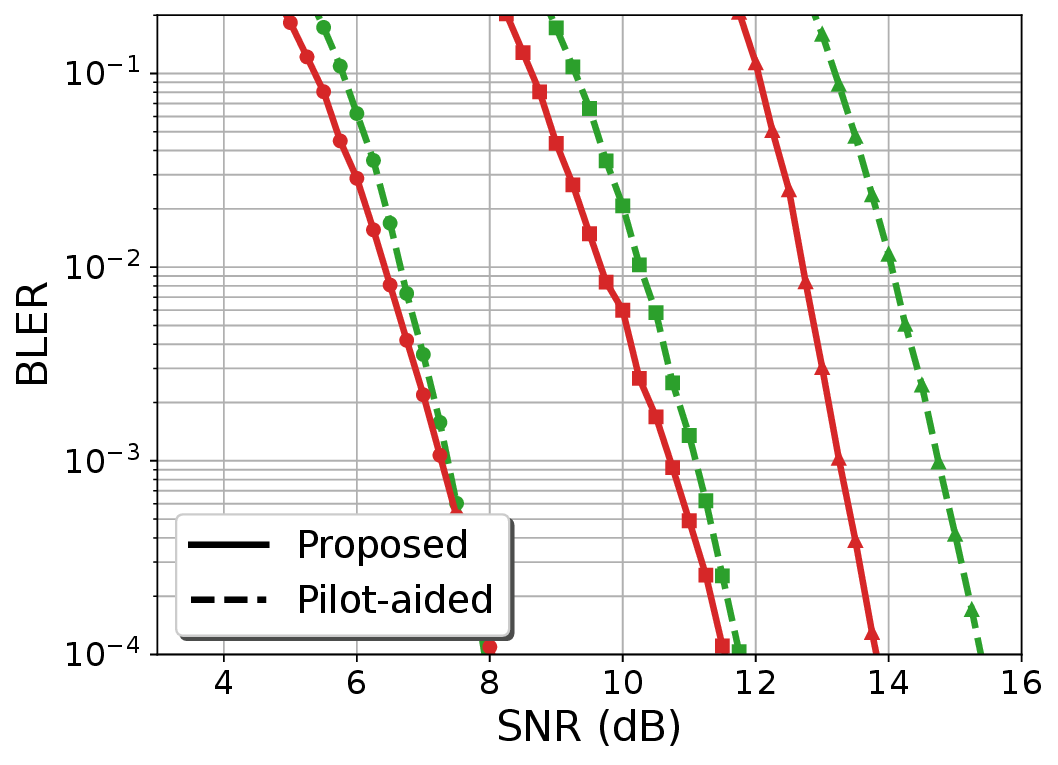}
    \hfil
    \includegraphics[width=0.32\textwidth]{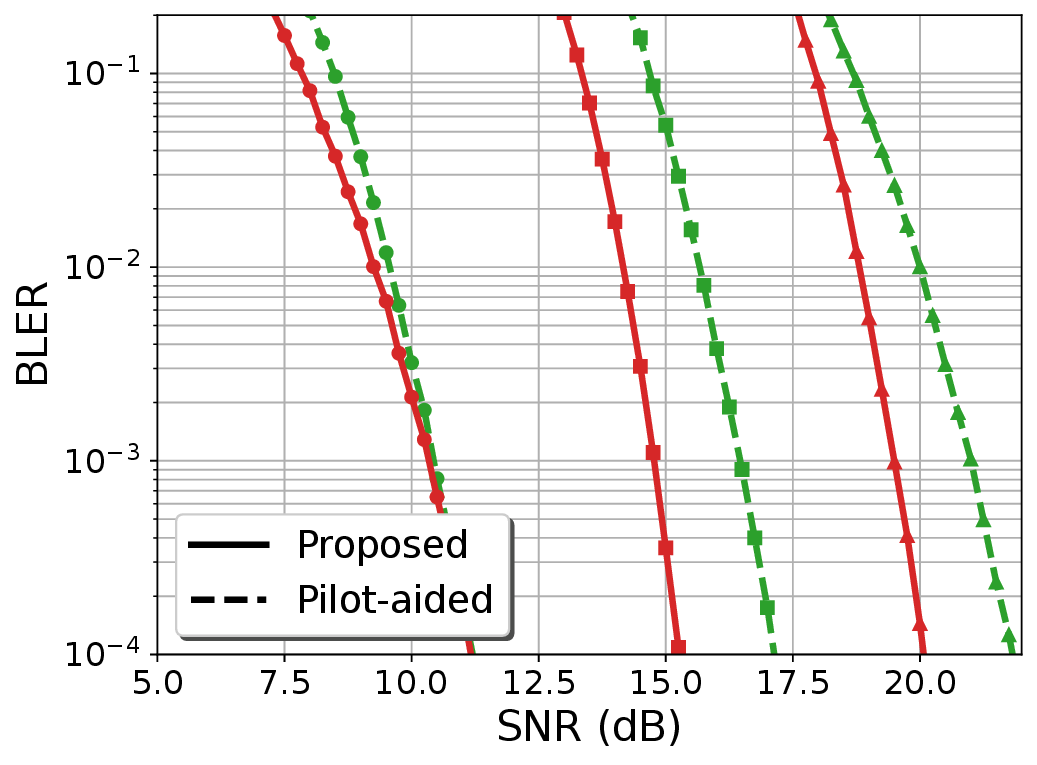}

    \parbox{0.32\textwidth}{\centering \footnotesize (a) 4-QAM}
    \hfil
    \parbox{0.32\textwidth}{\centering \footnotesize (b) 16-QAM}
    \hfil
    \parbox{0.32\textwidth}{\centering \footnotesize (c) 64-QAM}
    
    \caption{BLER performance comparison for three block fading ($B=3$) with $M=720$ coded bits and message lengths $K \in \{180, 360, 540\}$. The proposed coded-pilot scheme maintains performance advantages over the pilot-aided baseline at longer blocklengths, demonstrating scalability across different packet sizes and modulation orders.}
    \label{fig:experiments_BLER_B3}
\end{figure*}

We compare the proposed coded-pilot scheme against the pilot-aided baseline using SCL decoding with list size $L=8$. Two configurations are evaluated in Fig.~\ref{fig:experiments_BLER_B1} and Fig.~\ref{fig:experiments_BLER_B3}: $M=240$ coded bits with $B=1$, and $M=720$ coded bits with $B=3$, respectively. For each total blocklength $M$, three message lengths are tested: $K \in \{60, 120, 180\}$ for $M=240$, and $K \in \{180, 360, 540\}$ for $M=720$, yielding effective code rates of approximately 0.25, 0.5, and 0.75.

The performance comparison reveals distinct trends depending on the number of fading blocks. For the three block-fading case ($B=3$, Fig.~\ref{fig:experiments_BLER_B3}), the proposed scheme consistently outperforms the pilot-aided baseline across all modulation orders and code rates. The gain is particularly pronounced at higher code rates. For instance, in the 16-QAM configuration at $K=540$, the proposed scheme achieves approximately 1.5 dB gain at BLER of $10^{-3}$. Similar substantial gains are observed in the 64-QAM configuration at $K=360$.

In contrast, for the single block-fading case ($B=1$, Fig.~\ref{fig:experiments_BLER_B1}), the proposed scheme demonstrates performance advantages primarily at medium to high code rates ($R \geq 0.5$). At the lowest rate ($R=0.25$, $K=60$), the pilot-aided baseline achieves comparable or slightly better performance. This suggests that with a single fading block, the overhead of splitting the code into two components becomes less justified when ample coding resources are available at low rates. However, at higher rates such as $K=120$ for 16-QAM, the proposed scheme achieves approximately 1 dB gain at BLER of $10^{-3}$, demonstrating clear benefits when coding resources become constrained.

The performance improvement in favorable regimes stems from two factors: (i) the coded-pilot scheme eliminates dedicated pilot overhead by embedding channel estimation information within the codeword, and (ii) the joint optimization of code parameters $(M_0, K_0)$ and $(M_1, K_1)$ balances detection reliability and coding efficiency. The benefits are more pronounced in multi-block fading ($B=3$) where channel estimation information can be exploited across multiple coherence intervals, and at higher code rates where spectral efficiency becomes critical.


    

\subsection{Rate Allocation Strategy Analysis}

\begin{figure*}[t]
    \centering
    \includegraphics[width=0.32\textwidth]{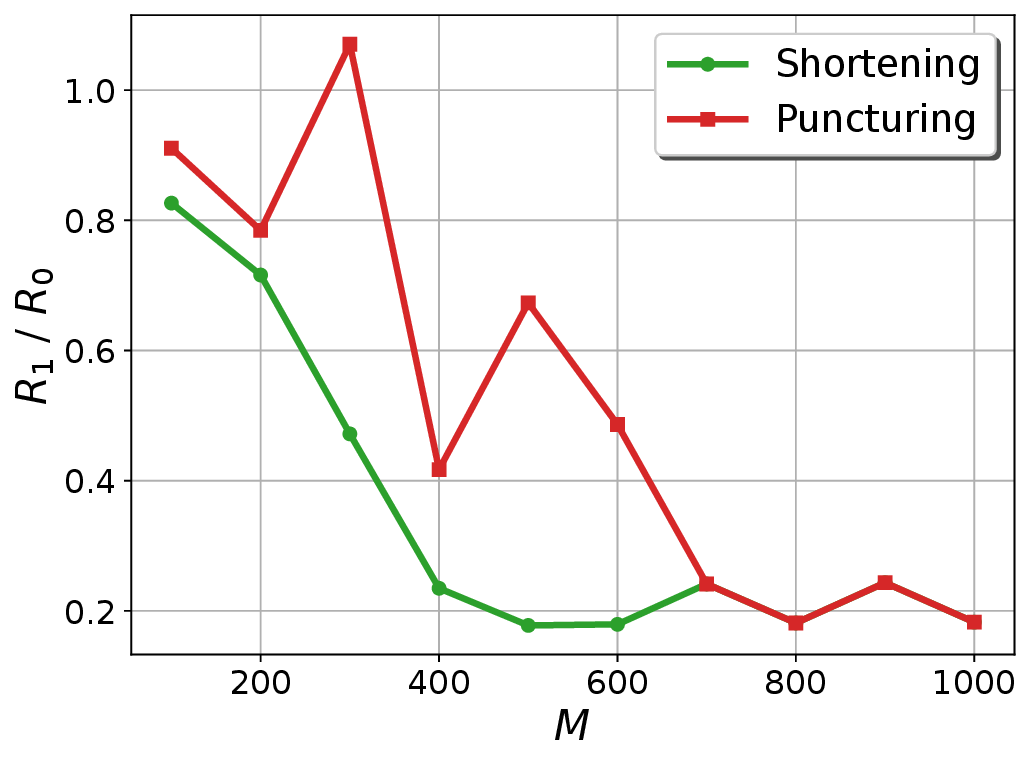}
    \hfil
    \includegraphics[width=0.32\textwidth]{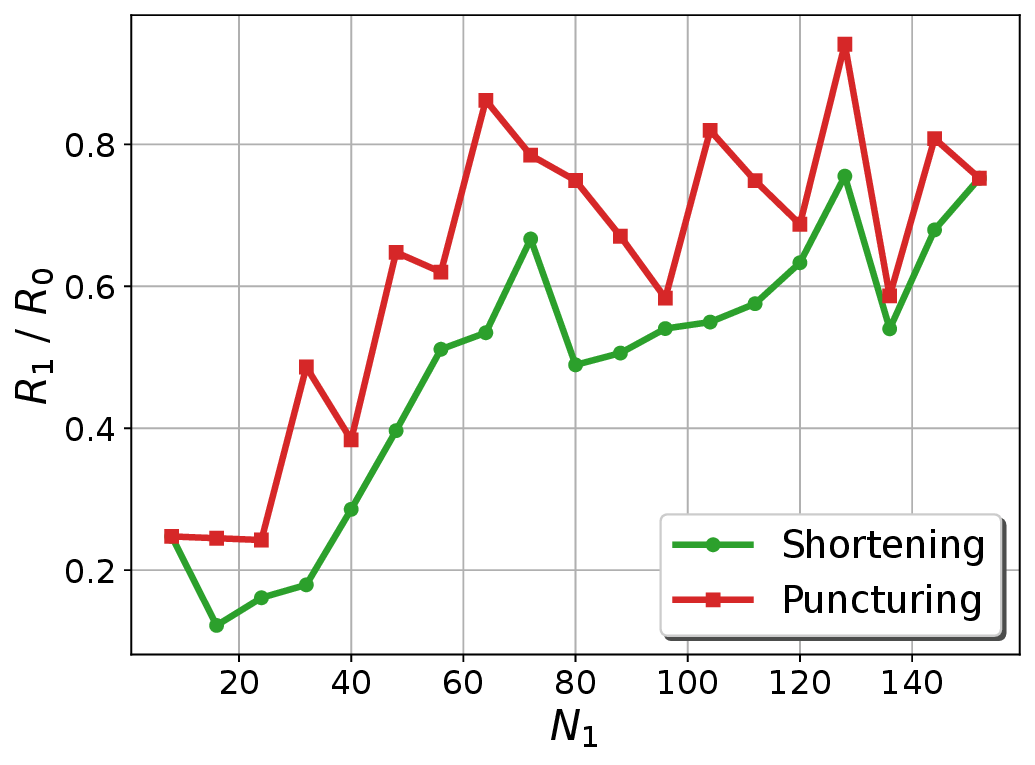}
    \hfil
    \includegraphics[width=0.32\textwidth]{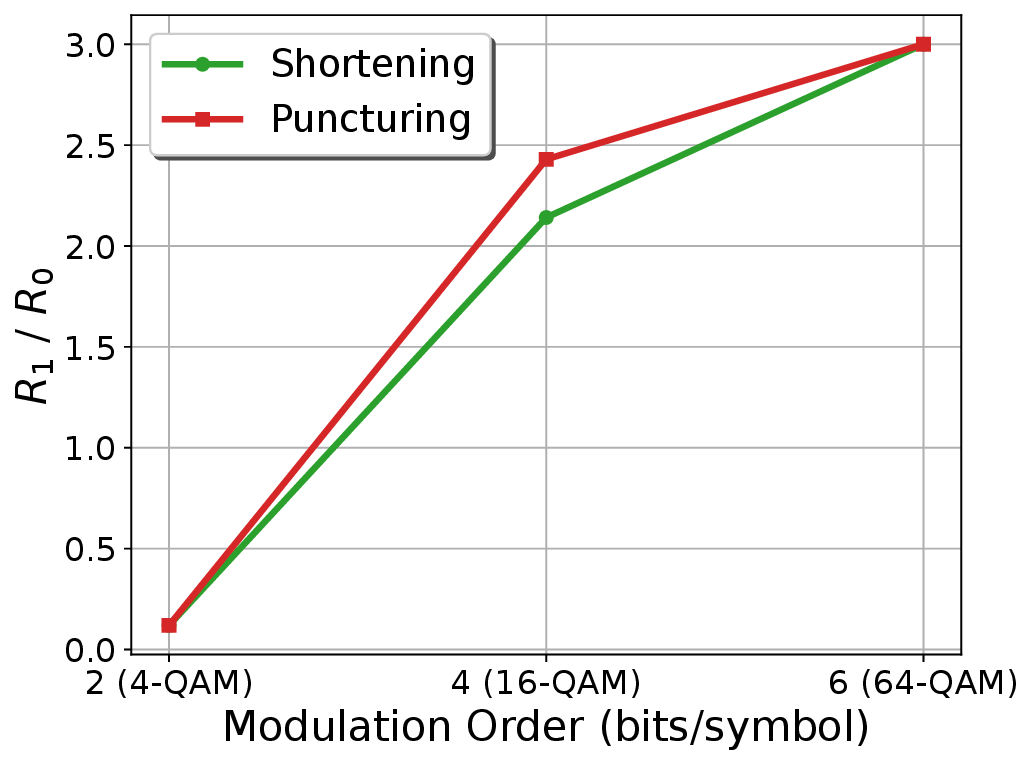}
    
    
    \parbox{0.32\textwidth}{\centering \footnotesize (a) Varying $M$}
    \hfil
    \parbox{0.32\textwidth}{\centering \footnotesize (b) Varying $N_1$}
    \hfil
    \parbox{0.32\textwidth}{\centering \footnotesize (c) Varying QAM order}
    
    \caption{Rate allocation ratio $R_1/R_0$ between coded-pilot and main data code as a function of system parameters. (a) Ratio decreases with total blocklength $M$ (fixed $M_1=32$, $K=0.5M$, 4-QAM). (b) Ratio increases with pilot code length $M_1$ (fixed $M=600$, $K=300$, 4-QAM). (c) Ratio increases with modulation order (fixed $M=600$, $K=150$, $M_1=32$). Green circles represent shortening and red squares represent puncturing for rate-matching of $\mathcal{C}_0$.}
    \label{fig:experiments_vary}
\end{figure*}




Fig.~\ref{fig:experiments_vary} illustrates how the rate allocation between the two component codes varies with system parameters. The rate ratio $R_1/R_0$ (where $R_1$ is the coded-pilot rate and $R_0$ is the main code rate) reveals distinct trends depending on  total blocklength $M$, coded-pilot length $M_1$, and modulation order.

Fig.~\ref{fig:experiments_vary}(a) shows the rate ratio as a function of total blocklength $M$. Increasing $M$ from 100 to 1000 causes $R_1$ to decrease while $R_0$ remains relatively stable. This indicates that channel estimation becomes the bottleneck for short packets, requiring higher pilot code rates. As $M$ grows, the system can afford lower $R_1$ while maintaining adequate detection performance. The rate ratio decreases from approximately 0.6 to 0.2, demonstrating that the effect of pilot symbols becomes much smaller at longer blocklength.

Fig.~\ref{fig:experiments_vary}(b) plots the rate ratio versus coded-pilot length $M_1$.
 As $M_1$ increases from 8 to 160, the pilot code rate $R_1$ increases for both puncturing and shortening. This reflects the need to improve pilot detection reliability when more resources are allocated to channel estimation. Simultaneously, $R_0$ decreases to compensate for the reduced main codeword length, preventing performance degradation in data detection. The rate ratio $R_1/R_0$ grows from approximately 0.2 to 0.8, showing that longer pilot codes require higher relative rates to maintain balanced error protection.

Fig.~\ref{fig:experiments_vary}(c) shows how the rate ratio varies with modulation order.
Higher-order modulation increases the rate ratio $R_1/R_0$. For 4-QAM, the ratio is approximately 0.1, while for 64-QAM it rises to 3. 
This trend arises because each pilot symbol occupies a channel use that could otherwise carry more data coded bits at higher modulation orders. For instance, in 64-QAM, each pilot symbol displaces 6 coded bits from the main code, compared to only 2 bits in 4-QAM. To offset this increased opportunity cost, $R_1$ must be raised accordingly.


\section{Conclusion}\label{sec:conclusion}

 This work shows that pilot overhead is not a law of nature—it is a design choice. By folding channel learning into the codeword itself, we avoid spending dedicated symbols on pilots while still acquiring the channel reliably. The key is the dual-packet structure architecture: a QPSK portion that simultaneously carries information and enables blind channel estimation, followed by a higher-order QAM portion that converts the recovered channel knowledge into spectral efficiency. This separation of roles—learn while you talk, then talk faster once you know—is what makes pilot-free operation practical without constraining modulation order or codeword length.

The net effect is simple: more of the packet is used for payload, not bookkeeping. In short packets, that accounting matters. Simulations confirm that reclaiming pilot symbols translates into a tangible performance gain—up to about a 1.5 dB coding advantage over conventional pilot-aided baselines—while preserving robust channel acquisition. The broader message is that, for ultra-reliable low-latency links, the right question is not how to optimize pilots, but how to design codes and modulation so that pilots become unnecessary.






\bibliographystyle{IEEEtran}
\bibliography{bibliography/IEEEAbbr, bibliography/coding}

\end{document}